\documentclass[12pt]{iopart}

\usepackage{amssymb}
\usepackage{cite}
\newcommand{\be}{\begin{equation}}
\newcommand{\ee}{\end{equation}}
\newcommand{\bea}{\begin{eqnarray}}
\newcommand{\eea}{\end{eqnarray}}
\newcommand{\hb}{\hbar}
\newcommand{\del}{\nabla^2}
\newcommand{\ka}{\kappa}
\newcommand{\qin}{(q^{-1})}
\newcommand{\dqin}{(\delta q^{-1})}
\newcommand{\etain}{(\bar{\eta}^{-1})}
\newcommand{\ebar}{\bar{\eta}}

\begin{document}

\title{The Parameterised Post-Newtonian Limit of Bimetric Theories of Gravity}

\author{Timothy Clifton$^{1}$, M\'{a}ximo
  Ba\~{n}ados$^{2}$\footnote{JS Guggenheim Memorial Foundation
    Fellow},\\ and Constantinos Skordis$^{3}$}

\address{
$^{1}$Department of Astrophysics, University of Oxford,
OX1 3RH, UK.\\
$^{2}$Departamento F\'{i}sica, P. Universidad Cat\'olica de Chile,
Santiago, Chile.\\
$^{3}$School of Physics and Astronomy, University of Nottingham,
NG7 2RD, UK.
}

\eads{\mailto{\mailto{t.clifton@cantab.net, \\ \hspace{31pt}
      maxbanados@gmail.com, \\ \hspace{31pt} skordis@nottingham.ac.uk
}}}


\begin{abstract}

We consider the post-Newtonian limit of a general class of
bimetric theories of gravity, in which both metrics are dynamical.
The established parameterised post-Newtonian approach is followed as
closely as possible, although new potentials are found that do not
exist within the standard frame-work.
It is found that these theories can evade solar system tests of
post-Newtonian gravity remarkably well.  We show that
perturbations about Minkowski space in these theories contain
both massless and massive degrees of freedom, and that in general there are
two different types of massive mode, each with a different mass parameter.  If both
of these masses are sufficiently large then the predictions of the
most general class of theories we consider are indistinguishable from
those of general relativity, up to post-Newtonian order in a
weak field, low velocity expansion.  In the limit that the massive modes become massless, we
find that these general theories do not exhibit a van Dam-Veltman-Zakharov-like discontinuity in their $\gamma$ parameter,
although there are discontinuities in other post-Newtonian parameters as the massless
limit is approached.  This smooth
behaviour in $\gamma$ is due to the discontinuities from each of the two different
massive modes cancelling each other out.  Such cancellations cannot occur in
special cases with only one massive mode, such as the Isham-Salam-Strathdee theory.

\end{abstract}

\maketitle

\section{Introduction}
\label{intro}

There exists a long history in gravitational physics of considering
theories that are generalisations of, or alternatives to, general
relativity.  These theories can take on a variety of different
forms, and often involve the introduction of extra scalar
\cite{scalar}, vector \cite{vector} or
tensor fields \cite{ISS} in the gravitational action.  Here we consider the
latter of these:  Bimetric theories in which there are two dynamical rank-2
symmetric tensor fields, rather than the usual one.  Our goal is to
calculate the Parameterised Post-Newtonian (PPN) limit of a general class of
these theories, so that their consequences for observational and
experimental gravitational physics can be determined in a
straightforward way.

The PPN formalism \cite{PPN1,PPN2} is now the standard frame-work within which
investigations of the phenomenology of relativistic gravitational
physics are performed \cite{PPN3}.  It has at its heart a simple and
compelling rationale: that by observationally constraining terms in a
general 'test metric', it is then possible to impose corresponding
constraints on a variety of different metric based gravitational
theories without going through the rigmarole of directly calculating
their physical consequences every time.
This allows the space-time geometry of the physical environments
in question to be constrained in a theory independent way, as well as
supplying a quick and direct route to determining the observational
validity (or otherwise) of particular theories.

The problem of determining the PPN limit of bimetric theories of
gravity has been approached in the past.  The PPN limit of the
bimetric theory of Rosen 
\cite{rosen} was found by Lee, Caves, Ni and Will
\cite{lee}.  The PPN limits of Rastall's bimetric theory
\cite{rastall1,rastall2}, and Lightman and Lee's bimetric theory
\cite{LL} are also known.  These theories, however, contain only
one dynamical rank-2 tensor each, the other being {\it a priori} specified
as being Riemann flat.  Such non-dynamical fields can cause serious
problems for these theories, particularly with respect to the
emission of gravitational waves from binary systems.  For example,
Rosen's theory predicts an unobserved {\it increase} in the rotational
period of the Hulse-Taylor binary pulsar system PSR 1913+16 \cite{WE},
and it can be 
argued that similar behaviour should be expected in all theories
with `priori geometry' \cite{PPN3}.  The bigravity theories considered in the
present article do not have any such non-dynamical fields, and are
therefore not expected to fall foul of binary pulsar observations in
the way discussed above.  To the best of our knowledge, the case of
determining the PPN limit of a general class of bimetric theories with
{\it two} dynamical rank-2 tensor fields has yet to be investigated in
detail.  It is this subject that we intend to investigate here.

As motivation for this study we use the recent astronomical
observations that have revealed that most of the matter in the
Universe is in the form of unknown ``dark components" (i.e. particles
or fields that are not present in the standard model,
and that do not interact electromagnetically).  Despite ongoing efforts
these fields have yet to be observed directly, and it seems natural to
consider extensions of general relativity that may explain them.
In this context,  bimetric gravity has recently been put
forward as an alternative to both dark matter and dark energy.  For
details of the way in which the extra degrees of freedom in these
theories can be made to reproduce the observations usually attributed
to the dark components of the Universe we refer the reader to
\cite{Damour:2002wu, B2, BFS, BGRS, SGGI, Milgrom, CZ}.  Here we will
not concern ourselves with the cosmological implications of these
theories directly, but will instead study their weak field limit.
If theories of this type are to be 
considered observationally viable it is absolutely necessary
that we properly understand their weak field limit, and how they are
constrained by observations of post-Newtonian gravitational phenomena.

The plan of this paper is as follows.  In Section \ref{bi} we
introduce and discuss the bimetric theories of gravity that we will be
investigating, giving their gravitational action and field
equations.  In Section \ref{PPNsec} we briefly introduce the PPN
formalism, the perturbative expansion it relies upon, and the way in
which it is used to constrain gravitational theories.  Section
\ref{newtsec} contains a calculation of metric perturbations up to
Newtonian accuracy.  We then proceed in Sections \ref{postnewtsec1}
and \ref{postnewtsec2} to derive the full post-Newtonian limit of the
metric perturbations about both of the metrics involved in these
theories.  In Section \ref{coupled} we transform into the standard
post-Newtonian gauge, and calculate the form of the perturbations in
the metric that couples to matter.  Section \ref{obs} provides a
discussion of the extent to which these theories can be subject to
observational constraint, and in Section \ref{disc} we provide a brief
discussion of our results.

\section{Bimetric Theories of Gravity}
\label{bi}

One of the most straightforward ways to extend general relativity is
to take the already existing concept of a dynamical rank-2
symmetric tensor field, and replace it with a multiplet of $N_g$ such fields.
Multigravity theories can then be constructed by considering an
extension of the Yang-Mills approach to gauge theories.  That is,
if we write our $N_g$ fields as $g_{\mu\nu}^{{a}}$, where $a=1,2,...,N_g$,
then our gravitational action can be written
\begin{equation}\label{naction}
S[g^{{a}}] \sim \int d^4 x \left[ \sum_{a=1}^{N_g} \sqrt{det(g^{{a}})}
  R(g^{{a}}) - I_{int}(g^{{a}}) \right],
\end{equation}
where $I_{int}$ is an interaction term that can depend on any or all of the
metrics.  Although it is impossible to construct a non-trivial $I_{int}$ term that
preserves the original $N_g$-dimensional symmetries of the free theory
\cite{Boulanger}, the action above is manifestly invariant
under diagonal diffeomorphisms acting on all metrics.  The potential
therefore breaks the symmetry group down to the diagonal subgroup of
diffeomorphisms acting on all metrics with the same parameter.
This approach to gravity has a long history, especially the $N_g=2$
incarnation known as `bigravity'. It began in the early 1970s with the pioneering
work of Isham, Salam and Strathdee  \cite{ISS}, while today actions like
(\ref{naction}) are mostly used as covariant actions for massive gravitons.
A classification of different interaction terms, as well as a variety
of different motivations for
bigravity, are explored in \cite{DamourKogan}.

In order to make progress we must specialise the general action
(\ref{naction}) somewhat.  The theory we will consider for the rest of
this paper is therefore the $N_g=2$ bigravity theory
\newpage
\bea
\hspace{-30pt}
S =\frac{1}{16 \pi G} \int d^4x \Bigg[ &\sqrt{-g} (R-2 \Lambda_0) +
  \sqrt{-q} (K-2 \lambda) \nonumber  \\&\qquad \qquad - \frac{1}{l^2} \sqrt{-q} \left( \ka_0 J + \ka_1
  J^2 + \ka_2 \qin^{\mu \nu} \qin_{\mu \nu} \right) \Bigg].
\label{general}
\eea
Here we have written the two metric fields as $g_{\mu \nu}$ and
$q_{\mu \nu}$.  The Ricci scalars constructed from these metrics are
written as $R$ and $K$, respectively, and the two constants
$\Lambda_0$ and $\lambda$ are the bare cosmological
constants on $g_{\mu \nu}$ and $q_{\mu \nu}$.  The three
constants $\ka_0$, $\ka_1$ and $\ka_2$ parameterise the interactions
between $g_{\mu \nu}$ and $q_{\mu \nu}$.  The inverse $\qin^{\mu \nu}$ is
defined such that $\qin^{\mu \alpha} q_{\mu \beta} = {\delta^{\alpha}}_{\beta}$, and
all raising or lowering of the indices of these tensors is done
with $g_{\mu \nu}$, so that e.g. $\qin_{\mu \nu} \equiv \qin^{\alpha \beta}
g_{\alpha \mu} g_{\beta \nu}$.  We have also defined $J \equiv \qin^{\mu \nu}
g_{\mu \nu}={\qin^{\mu}}_{\mu}$.

The theory above is a generalization of the often considered Isham-Salam-Strathdee (ISS) action
\be
S =\frac{1}{16 \pi G} \int d^4x \left[ \sqrt{-g} (R-2 \Lambda_0) +
  \sigma \sqrt{-f} (\hat{K}-2 \hat{\lambda}) - \frac{\sigma}{l^2}
  \sqrt{-f} I_{int}\right],
\label{pf}
\ee
with the interaction term
$I_{int}=(g_{\mu \nu} -f_{\mu \nu}) (g_{\alpha \beta}-f_{\alpha
    \beta}) ((f^{-1})^{\mu \alpha} (f^{-1})^{\nu \beta} -(f^{-1})^{\mu
  \nu} (f^{-1})^{\alpha \beta})$.
Here the scalar $\hat{K}$ is the Ricci scalar constructed from the
metric $f_{\mu \nu}$, and $(f^{-1})^{\mu \alpha} f_{\mu \beta} =
{\delta^{\alpha}}_{\beta}$.  The generalised theory (\ref{general}) can be seen to reduce to the ISS
action (\ref{pf})  under the metric rescaling $q_{\mu \nu} = \sigma f_{\mu \nu}$,
together with the identifications $\ka_0 =6$, $\ka_1=-\sigma$ and
$\ka_2=\sigma$, and the redefinition $\lambda= \hat{\lambda}/\sigma-6/(l^2 \sigma)$.

Of course, we also want to couple matter to these theories.  We do
this by including a Lagrangian density of the form
$\mathcal{L}_m=\mathcal{L}_m(\psi,\hat{g}_{\mu \nu})$, where the
matter fields $\psi$ are coupled to a linear combination of the
metrics
\be
\hat{g}_{\mu \nu}=m g_{\mu \nu}+n q_{\mu \nu},
\ee
where $m$ and $n$ are constants.  After
varying with respect to each of the metric fields this then results
in the field equations
\bea
 &&l^2 R_{\mu \nu} - 8 \pi G l^2 m \sqrt{\frac{\hat{g}}{g}} \left( T_{\mu \nu}-\frac{1}{2}
g_{\mu \nu} {T^{\alpha}}_{\alpha} \right) -
\alpha_0 g_{\mu \nu} \label{R} \\  &=&- \sqrt{\frac{q}{g}} \Bigg[ (\ka_0 +2
  \ka_1 J) \left( \qin_{\mu \nu}-\frac{1}{2}g_{\mu \nu} J \right)
  \nonumber \\ && 
+2 \ka_2 \left( \qin_{\mu \alpha} {\qin^\alpha}_{\nu} - \frac{1}{2}
  \qin_{\alpha \beta} \qin^{\alpha \beta} g_{\mu \nu} \right) \Bigg],
\nonumber
\eea
and
\bea
\label{K}
&& l^2 K_{\mu \nu} - 8 \pi G l^2 n \sqrt{\frac{\hat{g}}{q}}\left(q_{\mu \alpha} q_{\nu \beta} T^{\alpha \beta}
- \frac{1}{2} q_{\mu \nu} T^{\alpha \beta} q_{\alpha \beta}
\right) - \alpha q_{\mu \nu}\\
&=& -\frac{1}{2} \left[ \ka_1 J^2+\ka_2 \qin_{\alpha \beta}
  \qin^{\alpha \beta} \right] q_{\mu \nu} + \left[ \ka_0+2 \ka_1 J
  \right] g_{\mu \nu} +2 \ka_2 \qin_{\mu \nu},
\nonumber
\eea
where $\alpha_0\equiv \Lambda_0 l^2$ and $\alpha \equiv \lambda l^2$,
and where $T^{\mu \nu}$ is defined by
\be
T^{\mu \nu} =-\frac{2}{\sqrt{-\hat{g}}} \frac{\delta
  \mathcal{L}_m}{\delta \hat{g}_{\mu\nu}}.
\ee
Expressing the field equations in this form, without $R$ and $K$, the perturbation
equations take on their simplest form.

Now, we should be aware that these theories contain within their
spectrum a massive spin-$2$ field, and that such terms often have
associated with them certain discontinuities and instabilities
\cite{Blas2006,BlasDeffayetGarriga,Blas:2008uz,Berezhiani:2007zf,Berezhiani:2008nr,Berezhiani:2009kv,Blas:2009my}.
These include the van Dam-Veltman-Zakharov (vDVZ)
discontinuity~\cite{vanDamVeltman1970,Zakharov1970}, and the Boulware-Deser
(BD) instability \cite{bd}.  Both of these are potentially very serious
problems.  The vDVZ discontinuity involves the zero-mass limit of massive
theories of gravity.  Instead of recovering the solutions of general
relativity in this limit, it can instead sometimes
be found that space-time geometry approaches a limit which is
inconsistent with observed gravitational phenomena.  For example, the
prediction for weak lensing by the Sun in the Pauli-Fierz theory
\cite{PF} yields results which disagree with general relativity by
25\%.  The BD instability involves the possible existence
of negative energy states at the non-linear level, and a lack of
boundedness from below. We refer the reader to \cite{mass1, mass2,
  MuC} for recent discussions on massive (single metric) gravitons.

While deserving of careful study, these problems are not necessarily
fatal for the bimetric theories we are considering.  The vDVZ
discontinuity and the BD instability were both identified in the
context of bimetric theories in which only one of the metrics is
dynamical (the other being {\it a priori} specified as Riemann flat).
It remains to be seen to what extent these problems endure when both
tensor fields are dynamical.  Furthermore, in the case of the vDVZ
discontinuity it is thought that in the zero-mass limit the
graviton can become strongly-coupled, such that the usual perturbative analysis
can break down on scales smaller than the Vainshtein radius, so that general relativity is recovered \cite{Vain,AHGS}.  This `Vainshtein
mechanism' can be demonstrated numerically \cite{Babichev:2009jt}.
Also, in the theories we are considering the vDVZ discontinuity is not
necessarily a problem as both massless and massive modes can be shown
to exist when Minkowski space is perturbed.  The mass parameters
can therefore be assumed to be large, and there will still always exist long-ranged
massless modes to carry the Newtonian gravitational force (as long as
matter doesn't couple to the massive modes only).  The relevant
question in this case then becomes whether or not general relativity can be recovered
in the limit $M \rightarrow \infty$, rather than $M \rightarrow 0$.
This is primarily the question we will concern ourselves with here, although we
will also pause in the final sections to consider the zero-mass limit
of the theory.

\section{The PPN Approach}
\label{PPNsec}

Here we will briefly recap the essential elements of the PPN
formalism, as required for the internal coherence of this article. The PPN formalism is a
perturbative treatment, and requires a small parameter to expand in.  An ``order
of smallness'' is therefore defined by
\be
U \sim v^2\sim\frac{p}{\rho} \sim \Pi \sim O(2)
\ee
where $U$ is the Newtonian potential, $v$ is the velocity of a fluid
element, $p$ is the fluid pressure, $\rho$ is its rest-mass
density and $\Pi$ is the ratio of energy to rest-mass
densities.  We also take time derivatives to add an extra order of
smallness each:
\be
\frac{\partial}{\partial t} \sim O(1).
\ee
The PPN formalism then proceeds using an expansion in this order of smallness.
For our theories, we will now perturb our two fundamental tensors as
\bea
\label{gandq1}
g_{\mu \nu} &=& \eta_{\mu \nu} + h_{\mu \nu}\\
\label{gandq2}
q_{\mu \nu} &=& \ebar_{\mu \nu} + \hb_{\mu \nu},
\eea
where $\eta_{\mu \nu}$ is the metric of Minkowski space, and where we take
$
\ebar_{\mu \nu}=X_0^2 \eta_{\mu \nu},
$
where $X_0$ is a constant.  This last equation expresses the fact
that, in general, one set of coordinates cannot be used to correspond to the same
proper separation between events in the geometries associated with
each of the two metrics.  To be fully general one could consider
different constants in front of the different components of $\eta_{\mu \nu}$.  We will not
do this here.

Let us note that Minkowski space is not always a solution to the
field equations (\ref{R}) and (\ref{K}), due to constant terms in the
action.  For $ q_{\mu \nu}=X_0^2 g_{\mu \nu}$ these terms
correspond to the `cosmological constants' terms
\bea
 \rho_\Lambda = \frac{1}{8\pi G \ell^2}\left( \alpha_0  + X_0^2
 \kappa_0 + 8\kappa_1 + 2 \kappa_2\right)\\
\rho_\lambda = \frac{1}{8\pi G\ell^2} \left( \alpha  + \frac{\kappa_0}{X_0^2} \right),
\eea
for $g$ and $q$, respectively.  These can, in general, be non-zero even in
the absence of $\alpha$ and $\alpha_0$, resulting in a de Sitter background\footnote{An exception is the ISS
  theory, for which $ \rho_\Lambda = \alpha_0/8\pi G \ell^2$.}.  For vanishingly small
$\rho_\Lambda$ and $\rho_\lambda$, however, we can approximate this as
Minkowski space.  This corresponds to the conditions
\bea
\label{a0b}
\alpha_0 &\rightarrow&-X_0^2 \ka_0-8 \ka_1-2 \ka_2\\
\label{ab}
\alpha &\rightarrow& -\frac{\ka_0}{X_0^2},
\eea
which we will use to eliminate $\alpha_0$ and $\alpha$ in
the perturbed equations that follow.

Now consider the energy-momentum tensor, $T^{\mu \nu}$.  For a perfect
fluid this is given by
$
T^{\mu \nu} = \left[ \rho (1+\Pi)+p \right] u^{\mu} u^{\nu} +p \hat{g}^{\mu \nu},
$
where $\rho$ is the density of rest mass, $p$ is pressure, $\Pi$ is internal
energy per unit rest mass, and $u^{\mu}$ is the 4-velocity of matter.
Neglecting $O(1)$ contributions from $h_{0i}$ and $\hb_{0i}$,
the time-like normalisation $u^{\mu}u^{\nu} \hat{g}_{\mu \nu} =-1$ then gives
\be
u^{\mu} = \frac{1}{\sqrt{m+nX_0^2}} \left( 1+\frac{1}{2}v^2+ \frac{(m
  h_{00} + n \hb_{00})}{2 (m+n X_0^2)} +O(4) ;v^i  +O(3) \right),
\ee
to the relevant order in perturbations the components of the
energy-momentum tensor are then given by
\bea
T^{00} &=& \frac{\rho }{(m+n X_0^2)}\left[ 1+\Pi +v^2 +\frac{(m
      h_{00}+n\hb_{00})}{(m+nX_0^2)}  \right] +O(6) \label{em1}\\
T^{0i} &=& \frac{\rho v^i}{(m+n X_0^2)} +O(5) \label{em2}\\
T^{ij} &=& \frac{(\rho v^i v^j+p \delta^{ij})}{(m+n X_0^2)} +O(6) \label{em3}.
\eea
The two expanded metrics (\ref{gandq1}) and (\ref{gandq2}) can now be
substituted into the field equations (\ref{R}) and (\ref{K}), along
with (\ref{em1}), (\ref{em2}) and (\ref{em3}).
The field equations can then be solved for order by
order in smallness of perturbations, and gauge transformations of the form
\be
x^{\mu} \rightarrow x^{\mu} + \xi^{\mu}
\ee
can be used to transform it into the ``standard post-Newtonian gauge'', where the spatial part of
the metric is diagonal, and where terms containing time derivatives are
removed wherever possible.  The metrics that result then allow us to
determine the post-Newtonian parameters of these theories by
comparing to the test metric
\bea
\hspace{-30pt}\hat{g}_{00} = -1 +2 G U-2 \beta G^2 U^2-2 \xi G^2 \Phi_W +(2 \gamma +2 +
\alpha_3 +\zeta_1 -2 \xi) G \Phi_1 \nonumber \\
 +2 (1+3 \gamma -2 \beta +\zeta_2 +\xi) G^2 \Phi_2 +2 (1+\zeta_3)
G \Phi_3 \nonumber \\
 +2 (3 \gamma+3 \zeta_4-2 \xi) G \Phi_4 -(\zeta_1-2 \xi) G \mathcal{A} \label{pnm1}\\
\hspace{-30pt}\hat{g}_{0i} = -\frac{1}{2} (3+4 \gamma +\alpha_1-\alpha_2
+\zeta_1-2 \xi) G V_i -\frac{1}{2}(1+\alpha_2-\zeta_1+2 \xi) G W_i \label{pnm2}\\
\hspace{-30pt} \hat{g} _{ij} = (1+2 \gamma G U) \delta_{ij} \label{pnm3},
\eea
where $\beta$, $\gamma$, $\xi$, $\zeta_1$, $\zeta_2$, $\zeta_3$,
$\zeta_4$, $\alpha_1$, $\alpha_2$ and $\alpha_3$ are the
post-Newtonian parameters, $U$ is the Newtonian gravitational
potential that solves Poisson's equation, and $\Phi_W$, $\Phi_1$,
$\Phi_2$, $\Phi_3$, $\Phi_4$, $\mathcal{A}$, $V_i$ and $W_i$ are
the post-Newtonian gravitational potentials given in \cite{PPN3}.  The
particular combination of parameters before each of the
potentials in (\ref{pnm1}), (\ref{pnm2}) and (\ref{pnm3}) are given
so that the parameters have particular physical significance, once
gravitational phenomena have been computed.  We have chosen to
construct the test metric here in terms of the combined metric
$\hat{g}_{\mu \nu}$, as this is the metric that couples to the
matter fields, and hence is the one being constrained by observations
of those fields.

In general relativity the PPN parameters in (\ref{pnm1})-(\ref{pnm3})
are given by $\beta=\gamma=1$, with all other parameters equaling
zero.  The interpretation that is often given to these parameters is
that $\gamma$ is `the spatial curvature per unit rest mass', $\beta$
is `the degree of nonlinearity in the law of gravity', the $\alpha_i$
are due to `preferred-frame effects', and $\xi$ is due to
`preferred-location' effects.  The parameters $\zeta_i$, as well as
$\alpha_3$, are sometimes associated with the violation of
conservation of momentum.  These parameters are all constrained by
observations, and for the present study the two constraints below are
of particular interest \cite{gamma}:
\bea
\label{refgam}
\gamma -1 &=& (2.1 \pm 2.3) \times 10^{-5}\\
\vert \alpha_2 \vert &\lesssim& 1.2 \times 10^{-7}.
\label{refalph}
\eea
The first of these comes from observations of the Shapiro
time delay of radio signals from the Cassini space-craft as it passed
behind the Sun \cite{gamma}, and the second is derived from the
observation that the Sun's spin axis is closely aligned with the angular
momentum vector of the solar system \cite{a2}.  For further
constraints the reader is referred to \cite{PPN3}.

Finally, the equations of motion show that for time-like particles
propagating along geodesics the Newtonian limit is given by
$\hat{g}_{00}$ (and hence $g_{00}$ and $q_{00}$) to $O(2)$ and
$\hat{g}_{0i}$ to $O(1)$, with no other knowledge of the metric
components beyond the background level being necessary\footnote{The
  $O(1)$ terms in $\hat{g}_{0i}$ are usually omitted from the
  beginning in theories with a single metric, as they can be removed
  by a suitable gauge choice.  In what follows we will find that the
  situation is somewhat more complicated in multigravity.}.  The
post-Newtonian limit for time-like trajectories requires a knowledge of
\bea
&\hat{g}_{00} \qquad \textrm{to} \qquad O(4)\\
&\hat{g}_{0i} \qquad \textrm{to} \qquad O(3)\\
&\hat{g}_{ij} \qquad \textrm{to} \qquad O(2),
\eea
where Latin letters are used to denote spatial indices.  To obtain the
Newtonian limit of trajectories followed by null particles we only need to
know the metric to background order. The post-Newtonian limit of null trajectories
requires $\hat{g}_{00}$ and $\hat{g}_{ij}$ both to $O(2)$, as well as
$\hat{g}_{0i}$ to $O(1)$.

In order to proceed in calculating the form of the weak field metric
$\hat{g}_{\mu \nu}$, to the orders of smallness specified above, it is
useful to have expressions for various perturbed quantities involving
the metric tensors, such as their determinants and inverses.  These
are given in Appendix A.

\section{Newtonian Perturbations to {\boldmath $O(1)$} and {\boldmath $O(2)$}}
\label{newtsec}

\subsection{{\bf The {\boldmath $g_{0i}$} and {\boldmath $q_{0i}$} terms, to {\boldmath $O(1)$}}}

Unlike the case of gravitational theories with a single metric,
for the theories being considered here we cannot automatically set
both $g_{0i}=0$ and $q_{0i}=0$ to $O(1)$ through gauge transformations.  We must
therefore calculate these terms explicitly.

Substituting the expressions derived above into the $0-i$ components
of the field equations (\ref{R}) and (\ref{K}), and eliminating
$\alpha_0$ and $\alpha$ using (\ref{a0b}) and (\ref{ab}), gives
\bea
\label{R0i}
l^2 \left[ -\frac{1}{2} \del h_{0i} + \frac{1}{2}
  h_{0j,ji} \right]
&=& \left[ \ka_0+\frac{8 \ka_1}{X_0^2}+\frac{4 \ka_2}{X_0^2} \right] \left[ \hb_{0i}-
  X_0^2 h_{0i} \right]
\\
\label{K0i}
l^2 \left[ -\frac{1}{2} \del \hb_{0i} + \frac{1}{2}
  \hb_{0j,ji} \right]
&=& \left[ \ka_0+\frac{8 \ka_1}{X_0^2}+\frac{4 \ka_2}{X_0^2} \right]
\left[ X_0^2 h_{0i}- \hb_{0i} \right].
\eea
The left-hand side of these equations vanishes when acted upon with $\nabla \cdot$, due to the
Bianchi identities.  We must therefore have
$
\hb_{0i,i}= X_0^2 h_{0i,i}.
$
The equations (\ref{R0i}) and (\ref{K0i}) can then be decoupled by
  defining two new variables:
\newpage
\bea
h_{0i}^{(m)} &\equiv& h_{0i} - \frac{\hb_{0i}}{X_0^2}\\
h^{(0)}_{0i} &\equiv& h_{0i} + \hb_{0i},
\eea
to get
\bea
\del h^{(m)}_{0i} &=& M^2 h^{(m)}_{0i}
\\
\del h^{(0)}_{0i} &=& h^{(0)}_{0j,ji},
\eea
where we have defined
\be
\label{M}
M^2 \equiv \frac{2 (1+X_0^2)}{X_0^2 l^2} [X_0^2 \ka_0 +8 \ka_1 +4 \ka_2].
\ee
We can now set $h^{(0)}_{0i}=0$ by a gauge transformation; the
  $h^{(m)}_{0j}$ mode, however, is gauge invariant, and so cannot be
  made to vanish in this way. We now note that $
h^{(m)}_{0i,i}=0$, so $h^{(m)}_{0i}$ is the curl of
some vector potential $A_i$, with the solution
\be
h^{(m)}_{0i}=(\nabla \times A)_i = (c_1)_i \frac{e^{-M \vert x- x^{\prime} \vert}}{\vert x- x^{\prime} \vert} +
(c_2)_i \frac{e^{M \vert x- x^{\prime} \vert}}{\vert x- x^{\prime} \vert}.
\ee
If $M \in \mathcal{R}$ then $h^{(m)}_{0i}$ is a hyperbolic function, and if
$M \in \mathcal{I}$ then it is oscillatory.  This solution does not depend
on the matter content of the space-time, and exists in a vacuum where
only $g_{\mu \nu}$ and $q_{\mu \nu}$ are present.  It also has a
preferred point in space, $x^{\prime}$, and $h^{(m)}_{0i}$, being the
curl of a vector, looks like a rotation field.  Finally, we impose
$c_2=0$ as a boundary condition, in order to maintain asymptotic
flatness at infinity.  This is one of the PPN assumptions, up to
cosmological terms.

\subsection{{\bf The {\boldmath $g_{00}$} and {\boldmath $q_{00}$} terms, to {\boldmath $O(2)$}}}

We will now proceed to calculate $g_{00}$ and $q_{00}$ to $O(2)$.  To
do this we will need both the $0$-$0$ field equations to $O(2)$, and the trace of
the $i$-$j$ equations to $O(2)$.
%
As before, in order to decouple the equations let us now define massive and massless combinations
\bea
h^{(m)}_{\mu \nu} &\equiv& h_{\mu \nu}-\frac{\hb_{ \mu \nu}}{X_0^2}
\\
h^{(0)}_{\mu \nu} &\equiv& h_{\mu \nu}+\hb_{\mu \nu}.
\eea
With the gauge choice $h^{(0)}_{0i}=0$, and by using $h^{(m)}_{0i,i}=0$,
we can then write the relevant massless combination of (\ref{R}) and (\ref{K}) as
\bea
\nonumber
&& \del h^{(0)}_{00} + 8 \pi G (m+n X_0^2)^2 \rho \\
&=& \frac{X_0^2  }{(1+X_0^2)} \left( h^{(m)2}_{0i,j}-h^{(m)}_{0i,j}h^{(m)}_{0j,i}\right)
+ \frac{M^2 X_0^2}{2 (1+X_0^2)} h^{(m)2}_{0i},
\label{masslessNewton}
\eea
while the massive combination is given by
\newpage
\bea
\nonumber
&& \del h^{(m)}_{00} -M^2 h^{(m)}_{00} +8 \pi G (m+n X_0^2)(m-n) \rho
 \\ \nonumber
&=&  \frac{(1-X_0^2)}{(1+X_0^2)}\left(h^{(m)}_{0i,j}h^{(m)}_{0j,i}
-h^{(m)2}_{0i,j}\right) +\frac{2 (2 \ka_1+\ka_2) (1+X_0^2)
      }{l^2 X_0^2} A\\
&&+ \frac{(X_0^4 \ka_0 -8(1-X_0^2) \ka_1 -2(1-3X_0^2) \ka_2)}{l^2 X_0^2}h^{(m)2}_{0i}.
\label{massiveNewton}
\eea
Here we have defined
\be
A \equiv h^{(m)}_{ii} - h^{(m)}_{00},
\ee
which is given by the solution of
\bea
\nonumber
&& \left( X_0^2 \ka_0 +12 \ka_1+6 \ka_2 \right) \del
A - \ka_0 M^2 X_0^2 A
\\ \nonumber &=&
- \frac{8 \pi G X_0^2 M^2 l^2 (m+n X_0^2)(m-n)}{(1+X_0^2)} \rho
+ \frac{4 M^2 (8\ka_1+(3-X_0^2) \ka_2)}{(1+X_0^2)}
h^{(m)2}_{0i}
 \\ \nonumber
&& + \frac{(X_0^2 (1-X_0^2)\ka_0 +8 (3-X_0^2) \ka_1 +8 (1-X_0^2)
\ka_2)}{(1+X_0^2)} h^{(m)2}_{0i,j}\\
&&
+ \frac{(X_0^2 (1-X_0^2)\ka_0 +8 (1-X_0^2) \ka_1 +8 \ka_2)}{
  (1+X_0^2)} h^{(m)}_{0i,j} h^{(m)}_{0j,i}
.
\label{A}
\eea
To find the equations above we have made use of the expression $\del \left( h^{(m)2}_{0i} \right) = 2
h^{(m)2}_{0i,j}+2M^2 h^{(m)2}_{0i}$, as well as the Bianchi
identities to $O(2)$:
\bea
\hspace{-20pt}
\nonumber
\frac{M^2 l^2 X_0^2}{2 (1+X_0^2)} h^{(m)}_{ij,ij} &=& \frac{[X_0^2
    (1-X_0^2) \ka_0+8 (1-X_0^2) \ka_1+2 (3-X_0^2) \ka_2]}{(1+X_0^2)}
h^{(m)}_{0j,i} h^{(m)}_{0i,j}
\\&&
- \frac{[X_0^2 (1-X_0^2) \ka_0 -8 X_0^2 \ka_1 +2 (1-X_0^2)\ka_2]}{2
  (1+X_0^2)} \del \left( h^{(m)2}_{0i} \right)
\nonumber \\&&+
\frac{1}{2}
(X_0^2 \ka_0+4 \ka_1+2 \ka_2) \del \left( h^{(m)}_{ii}-h^{(m)}_{00}
\right)
.
\label{bianchi2}
\eea
It is interesting to note that in the ISS case the factor
multiplying the differential operator in (\ref{A}) vanishes.  The equation above
then becomes an algebraic relation between $A$, $\rho$ and
$h^{(m)}_{0i}$ (and its derivatives).  It can also be seen from
(\ref{A}) that in the case $\ka_0=0$ the Green's functions for $A$ are those
of Laplace's equation, rather than those of Helmholtz's equation.  In this
case $A$ becomes long-ranged.  In what
follows we will therefore consider the ISS and $\ka_0=0$ cases
separately from the more general case, as they clearly have different
behaviour.

We can now integrate Eq. (\ref{masslessNewton}) using the Green's
function of Laplace's equation.  The massless mode $h^{(0)}_{00}$ is
then given by
\bea
h^{(0)}_{00}
&=& 2 G (m+n X_0^2)^2U -\frac{M^2X_0^2}{8\pi (1+X_0^2)}\mathcal{V}\left(h^{(m)2}_{0i}\right)
\nonumber \\ && - \frac{X_0^2}{4 \pi (1+X_0^2)} \left[
  \mathcal{V}\left(h^{(m)2}_{0i,j}\right)-\mathcal{V}\left(h^{(m)}_{0i,j}h^{(m)}_{0j,i}\right) \right]
\eea
where the Newtonian potential $U$ is
\be
U = \int \frac{\rho(x^{\prime})}{\vert x-x^{\prime}
  \vert} d^3 x^{\prime}
\ee
and we have defined the additional functional
\be
\mathcal{V}(\phi) \equiv \int \frac{\phi(x^{\prime})}{\vert x-x^{\prime}
  \vert} d^3 x^{\prime}.
\ee
If $M$ is large, we are then left with the long-ranged part of this
metric component being given by
\be
h^{(0)}_{00} \dot{=} 2  G(m+n X_0^2)^2 U.
\ee
Here, and throughout, the symbol $\dot{=}$ will be taken to mean
`equal up to exponentially suppressed terms'.  Such suppression is
guaranteed as long as $M^{-1}$ is much smaller than the length scale
over which observations are being made, and can be arranged by choice of the
parameters $\kappa_0$, $\kappa_1$, $\kappa_2$, $l$ and $X_0$.

Let us now consider the massive modes.  These are different,
depending on the theory in question.

\vspace{15pt}
\noindent
\textbf{ISS theory}
\vspace{15pt}

For the ISS theory
we can write the solution to (\ref{massiveNewton}) and (\ref{A}) as
\bea
\nonumber
h^{(m)}_{00} &=&
\frac{8 G(m+n \sigma)(m-n)}{3}  \mathcal{W}_M \left(\rho\right)
+ \frac{(11-9 \sigma)}{24 \pi (1+\sigma)} \mathcal{W}_M \left(h^{(m)2}_{0i,j}\right)\\
&& - \frac{(1-3 \sigma)}{24 \pi (1+\sigma)} \mathcal{W}_M \left(h^{(m)}_{0i,j} h^{(m)}_{0j,i}\right)
+ \frac{(1-4 \sigma)}{6 \pi l^2} \mathcal{W}_M \left(h^{(m)2}_{0i}\right),
\eea
where we have defined the new functional
\be
\mathcal{W}_c(\phi) \equiv \int \frac{\phi(x^{\prime})}{\vert x-x^{\prime}
  \vert}e^{-c \vert x- x^{\prime} \vert} d^3 x^{\prime},
\ee
and suppressed the exponentially increasing mode.  For sufficiently
large $M$, the long-ranged part of this massive mode is zero:
\be
h^{(m)}_{00} \dot{=} 0.
\ee

\vspace{15pt}
\noindent
\textbf{$\mathbf{\ka_0=0}$ theory}
\vspace{15pt}

When $\ka_0=0$ the solution to equations (\ref{massiveNewton}) and (\ref{A}) is
\bea
\nonumber
h^{(m)}_{00}
&=& \frac{8 G}{3} (m+n X_0^2)(m-n) \mathcal{W}_M\left(\rho\right)
-\frac{A}{4}
\\&& \nonumber
+ \frac{(1-X_0^2)}{4
  \pi(1+X_0^2)}\left[\mathcal{W}_M\left(h^{(m)2}_{0i,j}\right)-\mathcal{W}_M\left(h^{(m)}_{0i,j}h^{(m)}_{0j,i}\right)\right]
\\&& + \frac{(4 (1-X_0^2)\ka_1+(1-3X_0^2) \ka_2)}{2\pi l^2 X_0^2} \mathcal{W}_M\left(h^{(m)2}_{0i}\right),
\eea

\newpage

\noindent
where
\bea
\nonumber
A&=&\frac{8 G (m+n X_0^2)(m-n)}{3} U
- \frac{((3-X_0^2) \ka_1+(1-X_0^2) \ka_2)}{3 \pi (1+X_0^2) (2\ka_1+\ka_2)}
\mathcal{V}\left(h^{(m)2}_{0i,j}\right)
\\ \nonumber && -\frac{((1-X_0^2) \ka_1+\ka_2)}{3 \pi (1+X_0^2) (2\ka_1+\ka_2)}
\mathcal{V}\left(h^{(m)}_{0i,j} h^{(m)}_{0j,i}\right)
\\ &&- \frac{4 (8\ka_1 +(3-X_0^2)\ka_2)}{3 \pi l^2 X_0^2}
\mathcal{V}\left(h^{(m)2}_{0i}\right),
\label{A2}
\eea
and where we have used $3 M^2 \mathcal{W}_M (A) = 12 \pi  A -32\pi G (m+n
X_0^2) (m-n)\mathcal{W}_M(\rho)$.  From the above it can be seen that
even for large $M$ the massive modes are not short-ranged:
\be
h^{(m)}_{00} \dot{=} -\frac{2 G}{3} (m+n X_0^2)(m-n) U.
\ee

\vspace{15pt}
\noindent
\textbf{Other theories}
\vspace{15pt}

For all other theories the solution to equations (\ref{massiveNewton}) and (\ref{A}) is
\bea
\nonumber
h^{(m)}_{00}
&=&
  \frac{8 G}{3} (m+n X_0^2)(m-n) \mathcal{W}_M\left(\rho\right)
- \frac{\ka_0 (1+X_0^2)}{3 l^2 N^2} A
\\&& \nonumber+ \frac{(1-X_0^2)}{4
  \pi(1+X_0^2)}\left[\mathcal{W}_M\left(h^{(m)2}_{0i,j}\right)-\mathcal{W}_M\left(h^{(m)}_{0i,j}h^{(m)}_{0j,i}\right)\right]
\\&& -\frac{(X_0^4 \ka_0-8 (1-X_0^2) \ka_1-2 (1-3X_0^2) \ka_2)}{4\pi
  l^2 X_0^2} \mathcal{W}_M\left(h^{(m)2}_{0i}\right),
\eea
where
\bea
\nonumber
   A
&=&
  \frac{2 G X_0^2 M^2 l^2 (m+n X_0^2)(m-n)}{(X_0^2 \ka_0 +12 \ka_1+6 \ka_2)(1+X_0^2)} \mathcal{W}_N\left(\rho \right)
\\&&\nonumber - \frac{(X_0^2 (1-X_0^2)\ka_0 +8 (3-X_0^2) \ka_1 +8
  (1-X_0^2)\ka_2)}{4 \pi (X_0^2 \ka_0 +12 \ka_1+6 \ka_2)(1+X_0^2)} \mathcal{W}_N\left(h^{(m)2}_{0i,j}\right)
\\&&\nonumber -\frac{(X_0^2 (1-X_0^2) \ka_0+8(1-X_0^2) \ka_1 +8\ka_2)}{4\pi (1+X_0^2) (X_0^2\ka_0+12\ka_1+6\ka_2)}
\mathcal{W}_N\left(h^{(m)}_{0i,j} h^{(m)}_{0j,i}\right)
\\&& \label{A3}-\frac{M^2 (8\ka_1+(3-X_0^2) \ka_2)}{\pi (1+X_0^2)
  (X_0^2 \ka_0+12\ka_1+6\ka_2)} \mathcal{W}_N\left(h^{(m)2}_{0i}\right),
\eea
and where we have defined
\be
\label{N}
N^2
\equiv \frac{2 \ka_0 (1+X_0^2) (X_0^2 \ka_0 +8
  \ka_1+4 \ka_2)}{l^2 (X_0^2 \ka_0 +12 \ka_1+6 \ka_2)},
\ee
and used
\be
\hspace{-35pt}
\frac{2 \ka_0 (1+X_0^2)}{l^2 N^2} A
= 4 G (m+n X_0^2)(m-n) \mathcal{W}_M(\rho) + \frac{3
  (2\ka_1+\ka_2)(1+X_0^2)}{\pi l^2 X_0^2} \mathcal{W}_M(A).
\ee
Now, if both $M$ and $N$ are large enough then there are no
long-ranged terms in the massive mode:
\be
h^{(m)}_{00} \dot{=} 0.
\ee

\section{Post-Newtonian Perturbations to {\boldmath $O(2)$} and {\boldmath $O(3)$}}
\label{postnewtsec1}

\subsection{{\bf The {\boldmath $g_{ij}$} and {\boldmath $q_{ij}$} terms, to {\boldmath $O(2)$}}}

Let us now consider the $g_{ij}$ and $q_{ij}$ modes to $O(2)$, which we call
`post-Newtonian tensor modes', as they are tensors under
transformations of the spatial coordinate system.
From now on we will not consider further the contribution of
the $h^{(m)}_{0i}\sim O(1)$ terms to the metric perturbations.  If they do not have
an effect at the Newtonian level, then it seems unlikely they will be
required at higher orders.


Now, if we apply the gauge condition
\be
\label{gauge2}
{h^{(0) \mu}}_{i ,\mu}- \frac{1}{2} {h^{(0) \mu}}_{\mu, i} =0
\ee
then the massless $i$-$j$ equation, to $O(2)$, becomes
\be
\nabla^2 h^{(0)}_{ij} = - 8 \pi G (m+n X_0^2)^2\delta_{ij} \rho,
\ee
which has the solution
\be
h^{(0)}_{ij} = 2 G (m+n X_0^2)^2\delta_{ij} U.
\ee
%
%
The massive modes, however, are once again dependent on the theory
that is being considered.  Now, the $i$-$j$ equations, to $O(2)$, are
\bea
\nonumber
&& \nabla^2 h^{(m)}_{ij} -M^2 h^{(m)}_{ij}+8 \pi G (m+n X_0^2)(m-n) \delta_{ij} \rho\\
&=&  - \frac{2
  (2\ka_1+\ka_2)}{(X_0^2 \ka_0 +8 \ka_1+4 \ka_2)} A_{,ij}-\frac{2
(2\ka_1+\ka_2) (1+X_0^2)}{l^2 X_0^2} \delta_{ij} A, \label{ij2}
\eea
where we have used the Bianchi identities to $O(2)$, Equation (\ref{bianchi2}),
and where $A$ is again given by the solution of Equation (\ref{A}).

\vspace{15pt}
\noindent
\textbf{ISS theory}
\vspace{15pt}

In the ISS theory the solution to (\ref{A}) and (\ref{ij2}) is
given by
\be
h^{(m)}_{ij} = -\frac{4 G}{3 M^2}(m+n \sigma)(m-n)
\mathcal{W}_M(\rho_{,ij}) +\frac{4 G}{3}(m+n \sigma)(m-n) \delta_{ij} \mathcal{W}_M(\rho).
\ee
The term involving $\mathcal{W}_M(\rho_{,ij})$ is not in keeping
with the usual PPN philosophy of avoiding terms with derivatives of
$\rho$ involved.  Normally, one could perform a gauge transformation
to remove such terms.  Here, the massive modes are gauge invariant,
but in subsequent sections we will find it possible to remove these
terms from the combination of metrics that couples to the matter fields.
For large $M$ we have
\be
h^{(m)}_{ij} \dot{=} 0.
\ee

\newpage

\vspace{15pt}
\noindent
\textbf{$\mathbf{\ka_0=0}$ theory}
\vspace{15pt}

When $\ka_0=0$ the solution to (\ref{A}) and (\ref{ij2}) is
\bea
\nonumber
h^{(m)}_{ij}
&=&  \frac{2 G}{3} (m+n X_0^2)(m-n)
\delta_{ij} U +\frac{4 G}{3} (m+n X_0^2)(m-n)
\delta_{ij} \mathcal{W}_M(\rho)
\\&&+\frac{G}{3\pi} (m+n X_0^2)(m-n)
\mathcal{W}_M(U_{,ij}).
\eea
There is again an unsightly term
containing derivatives of its integrand, which
we will gauge transform away later on when considering the combined
metric $\hat{g}_{ij}$.  The long-range component of $h^{(m)}_{ij}$ is
now given by
\be
h^{(m)}_{ij} \dot{=}\frac{2 G}{3} (m+n X_0^2)(m-n) \delta_{ij} U.
\ee

\vspace{15pt}
\noindent
\textbf{Other theories}
\vspace{15pt}

For all other theories the solutions to (\ref{A}) and (\ref{ij2}) are
given by
\bea
\nonumber \hspace{-20pt}
h^{(m)}_{ij}
&=& \frac{4 G}{3}(m+n X_0^2)(m-n) \delta_{ij} \mathcal{W}_M(\rho)+\frac{2 G}{3} (m+n X_0^2)(m-n) \delta_{ij} \mathcal{W}_N(\rho)
\\ && \qquad + \frac{(2\ka_1+\ka_2)}{2 \pi (X_0^2 \ka_0 +8 \ka_1+4 \ka_2)}
 \mathcal{W}_M(A_{,ij}),
\eea
where $A$ is given by (\ref{A3}).  Once again it can be seen that in
the general case there are two different massive modes, with masses $M$ and
$N$.  Once again, there is also a term containing derivatives.  If
both $M$ and $N$ are large enough then all of the terms in this
massive mode are suppressed:
\be
h^{(m)}_{ij} \dot{=} 0.
\ee

\subsection{{\bf The {\boldmath $g_{0i}$} and {\boldmath $q_{0i}$} terms, to {\boldmath $O(3)$}}}

Now consider the $g_{0i}$ and $q_{0i}$ modes to $O(3)$, which we call
`post-Newtonian vector modes', due to their properties under
spatial coordinate transformations.
If we again apply the gauge condition (\ref{gauge2}), together with
the new condition
\be
{h^{(0)\mu}}_{0,\mu}-\frac{1}{2} {h^{(0) \mu}}_{\mu,0} = -\frac{1}{2}
h^{(0)}_{00,0},
\label{gauge3}
\ee
then we can write the massless combination of the field equations
(\ref{R}) and (\ref{K}) as
\be
\label{nab0i}
\nabla^2 h^{(0)}_{0i} = \frac{G}{2} (m+n X_0^2)^2 \nabla^2 (V_i-W_i)
+16 \pi G(m+n X_0^2)^2 \rho v_i,
\ee
where we have used $U_{,0i}=\frac{1}{2} \nabla^2 (W_i-V_i)$, as well
as the previously found solution for $h^{(0)}_{00}$.  It can then be
seen that equation (\ref{nab0i}) has the solution
\be
h^{(0)}_{0i} = -\frac{7G}{2}(m+n X_0^2)^2 V_i - \frac{G}{2}(m+n X_0^2)^2 W_i,
\ee
where $V_i$ and $W_i$ are the usual vector post-Newtonian potentials,
as defined in \cite{PPN3}.

If we now use the $O(2)$ Bianchi identities, (\ref{bianchi2}), as well as
the Bianchi identities to $O(3)$,
\be
\hspace{-20pt}
\label{bianchi3}
2 (X_0^2\ka_0+8\ka_1+4\ka_2) h^{(m)}_{0i,i} = (X_0^2
 \ka_0+4\ka_1+2\ka_2) h^{(m)}_{ii,0}+
 (X_0^2\ka_0+12\ka_1+6\ka_2)h^{(m)}_{00,0},
\ee
then we can write the relevant combination of field equations from (\ref{R}) and (\ref{K})
as
\be
\hspace{-20pt}
\label{0i3}
\nabla^2 h^{(m)}_{0i} -M^2 h^{(m)}_{0i} = -\frac{2 (2
 \ka_1+\ka_2)}{(X_0^2 \ka_0+8\ka_1+4\ka_2)} A_{,0i} +16 \pi G
 (m+n X_0^2)(m-n) \rho v_i.
\ee
As above, the form of the massive modes in $\hat{g}_{0i}$ depend
on the theory being considered.

\vspace{15pt}
\noindent
\textbf{ISS theory}
\vspace{15pt}

The solution to Equation (\ref{0i3}) in the ISS theory is
given by
\be
h^{(m)}_{0i}= -\frac{Gl^2 (m+n \sigma)(m-n)}{3 (1+\sigma)} \mathcal{W}_M
(\rho_{,0i}) - 4 G(m+n \sigma)(m-n) \mathcal{W}_M (\rho v_i).
\ee
As with the $O(2)$ components, there is still only one mass, $M$.  If
$M$ is large enough then
\be
h^{(m)}_{0i}\dot{=} 0.
\ee

\vspace{15pt}
\noindent
\textbf{$\mathbf{\ka_0=0}$ theory}
\vspace{15pt}

When $\ka_0=0$ the solution to Equation (\ref{0i3}) is
given by
\bea
h^{(m)}_{0i}
&=& \frac{Gl^2 X_0^2 (m+nX_0^2)(m-n)}{6 (1+X_0^2) (2 \ka_1+\ka_2)}
U_{,0i}  - 4 G(m+n X_0^2)(m-n)
\mathcal{W}_M (\rho v_i) \nonumber \\&&
- \frac{Gl^2 X_0^2 (m+nX_0^2) (m-n)}{6 (1+X_0^2)
  (2\ka_1+\ka_2)} \mathcal{W}_M(\rho_{,0i}).
\eea
Possible long-ranged forces are now given by
\be
h^{(m)}_{0i}\dot{=} \frac{Gl^2 X_0^2 (m+nX_0^2)(m-n)}{6 (1+X_0^2) (2 \ka_1+\ka_2)}
U_{,0i}.
\ee


\vspace{15pt}
\noindent
\textbf{Other theories}
\vspace{15pt}

For all other theories we have that the solution to (\ref{0i3}) is
given by
\bea
h^{(m)}_{0i}
&=& \frac{2 \ka_0 (1+X_0^2)}{3 l^2 N^2 M^2} A_{,0i} - \frac{4G
  (m+nX_0^2)(m-n)}{3M^2} \mathcal{W}_M(\rho_{,0i}) \nonumber\\&&- 4G (m+n X_0^2)(m-n)
\mathcal{W}_M (\rho v_i) .
\eea
To $O(3)$ we have no long-ranged terms in $h^{(m)}_{0i}$, if $M$ and
  $N$ are both sufficiently large:
\be
h^{(m)}_{0i} \dot{=}0 .
\ee

\newpage

\section{Post-Newtonian Perturbations to {\boldmath $O(4)$}}
\label{postnewtsec2}

We will now consider the $g_{00}$ and $q_{00}$ component of the metric to $O(4)$.

\subsection{{\bf The {\boldmath $g^{(0)}_{00}$} and {\boldmath
      $q^{(0)}_{00}$} terms, to {\boldmath $O(4)$}}}

Using known solutions and identities we can write the massless
combination of the field equations (\ref{R}) and (\ref{K}) as
\bea
\hspace{-50pt}
  \nabla^2 h^{(0)}_{00}  = -8\pi G (m+nX_0^2)^2 \rho \left[1+\Pi+2
  v^2 +3\frac{p}{\rho}\right]
+ \frac{8 \pi G X_0^2 (m+nX_0^2)(m-n) \rho
  h^{(m)}_{00}}{(1+X_0^2)}
\nonumber \\ -\frac{h^{(0)2}_{00,i}}{(1+X_0^2)}
 -\frac{X_0^2 ( h^{(m)2}_{00,i}
  -h^{(m)}_{00,ij}h^{(m)}_{ij})}{(1+X_0^2)} -\frac{2
   (2\ka_1+\ka_2)}{M^2 l^2} h^{(m)}_{00,i}A_{,i}
\nonumber \\
 -\frac{X_0^2 \ka_0}{2 l^2} h^{(m)2}_{00}
-\frac{1}{4 l^2} (X_0^2\ka_0 +12 \ka_1+6 \ka_2) A^2
+\frac{M^2 X_0^2}{4 (1+X_0^2)} h^{(m)2}_{ij}.
\eea
This equation can be integrated to give
\bea
\hspace{-50pt}
  h^{(0)}_{00} = 2 (m+nX_0^2)^2 G\left[U
  -\frac{(m+nX_0^2)^2}{(1+X_0^2)} G U^2+2
  \Phi_1+ 2 \frac{(m+nX_0^2)^2}{(1+X_0^2)} G  \Phi_2 +\Phi_3 +3 \Phi_4 \right]
\nonumber \\ \hspace{-15pt}
- \frac{2 G X_0^2 (m+nX_0^2)(m-n) }{(1+X_0^2)} \mathcal{V} (\rho
  h^{(m)}_{00})+\frac{X_0^2 \ka_0}{8\pi l^2} \mathcal{V}(h^{(m)2}_{00})
\nonumber \\ \hspace{-15pt}
 +\frac{X_0^2 \left( \mathcal{V}(h^{(m)2}_{00,i})
  -\mathcal{V}(h^{(m)}_{00,ij}h^{(m)}_{ij})\right)}{4 \pi(1+X_0^2)}
+\frac{(2\ka_1+\ka_2)}{2 \pi M^2 l^2} \mathcal{V}(h^{(m)}_{00,i}A_{,i})
\nonumber \\ \hspace{-15pt}
+\frac{(X_0^2\ka_0 +12 \ka_1+6 \ka_2)}{16\pi l^2} \mathcal{V}(A^2)
-\frac{M^2 X_0^2}{16\pi (1+X_0^2)} \mathcal{V}(h^{(m)2}_{ij}),
\eea
where the $h^{(m)}_{00}$,  $h^{(m)}_{ij}$ and $A$ to $O(2)$ are taken
  to be given by the expressions found in previous sections.  The
  long-ranged modes in $h^{(0)}_{00}$ are then given, for our various
  different theories by the following:


\bigskip
\noindent
{\bf ISS theory}
\vspace{10pt}
\bea
\hspace{-30pt}
  h^{(0)}_{00}
\dot{=}
2 (m+n\sigma)^2 G \left[U
  -\frac{(m+n\sigma)^2}{(1+\sigma)} G U^2+2
  \Phi_1+ 2 \frac{(m+n\sigma)^2}{(1+\sigma)} G  \Phi_2 +\Phi_3 +3 \Phi_4
  \right]
\nonumber \\
+\frac{4 \pi \sigma G^2 l^2 (m+n\sigma)^2(m-n)^2}{3
  (1+\sigma)^2}
\mathcal{V}(\rho^2).
\eea

\newpage

\bigskip
\noindent
{\bf $\mathbf{\ka_0=0}$ theory}
\vspace{10pt}
\bea
\hspace{-30pt}
h^{(0)}_{00} \dot{=} 2 (m+nX_0^2)^2 G \left[U
  -\frac{(m+nX_0^2)^2}{(1+X_0^2)} G U^2+2
  \Phi_1+ 2 \frac{(m+nX_0^2)^2}{(1+X_0^2)} G  \Phi_2 +\Phi_3 +3 \Phi_4
  \right]
\nonumber \\
+ \frac{8 X_0^2 G^2 (m+nX_0^2)^2(m-n)^2}{3 (1+X_0^2)} \Phi_2
+\frac{M^2X_0^2G^2 (m+nX_0^2)^2(m-n)^2}{4 \pi (1+X_0^2)}\mathcal{V}(U^2)
\nonumber \\
-\frac{2X_0^2G^2 (m+nX_0^2)^2 (m-n)^2}{9(1+X_0^2) M^2} \vert \nabla U \vert^2
-\frac{4 X_0^2G^2 (m+nX_0^2)^2(m-n)^2}{3 M^2
  (1+X_0^2)}\mathcal{V}(\rho_{,i} U_{,i})
\nonumber \\
+\frac{4 X_0^2G^2 (m+nX_0^2)^2(m-n)^2}{3 M^2
  (1+X_0^2)}\mathcal{V}(\rho_{,i} \mathcal{W}_M(\rho)_{,i}).
\eea

\bigskip
\noindent
{\bf Other theories}
\vspace{10pt}
\bea
\hspace{-60pt}
  h^{(0)}_{00} 
\dot{=} 2 (m+nX_0^2)^2 G \Bigg[U
  -\frac{(m+nX_0^2)^2}{(1+X_0^2)} G U^2+2
  \Phi_1
+ 2 \frac{(m+nX_0^2)^2}{(1+X_0^2)} G  \Phi_2 +\Phi_3 +3 \Phi_4
  \Bigg].\hspace{30pt}
\eea

\bigskip
\noindent
In deriving these expressions we have used $\vert \nabla U \vert^2 =
\frac{1}{2} \nabla^2 U^2-\nabla^2 \Phi_2$, as well as
\bea
A_{,i} B_{,i} &=& \frac{1}{2} \nabla^2 (A B) +2 \pi \sigma A +2
\pi \rho B-\frac{1}{2}(M^2+N^2)A B \label{id1}\\
\nonumber A_{,ij} B_{,ij} &=& \frac{1}{2} \nabla^2 (A_{,i}B_{,i})+2 \pi
\sigma_{,i} A_{,i}+2 \pi \rho_{,i} B_{,i} -\frac{1}{4} (M^2+N^2)
\nabla^2 (AB)
\\&&\qquad+\frac{1}{4}
(M^2+N^2)^2 AB-\pi (M^2+N^2)(A\sigma + B \rho )
\label{id2}
\eea
where $A=\mathcal{W}_M(\rho)$ and $B=\mathcal{W}_N(\sigma)$.
Further manipulation has also been done by noting that, for example,
$\mathcal{V}(\rho_{,i} \mathcal{W}_M(\rho)_{,i}) \dot{=} 4 \pi
\mathcal{V}(\rho^2)$.  These manipulations are performed so that the potentials are
written without derivatives acting on the matter fields, as is usual
in the PPN formalism.

\subsection{{\bf The {\boldmath $g^{(m)}_{00}$} and {\boldmath $q^{(m)}_{00}$} terms, to {\boldmath $O(4)$}}}

The equations involved in finding $g^{(m)}_{00}$ and $q^{(m)}_{00}$ to $O(4)$ are
very lengthy, and we therefore choose to present them in Appendix B.
Here we only state the form of the equations, and give their
long-ranged components when $M$ and $N$ are large.  We find the equation for $h^{(m)}_{00}$ looks
like
\be
\label{hm00}
\nabla^2 h^{(m)}_{00} -M^2 h^{(m)}_{00} =... .
\ee
The Green's function for this equation is therefore that of
Helmholtz's equation, with the same mass term, $M$, as at $O(2)$.
The right-hand side of this equation has terms containing $A$ to
$O(4)$.  This quantity is given in Appendix B by the solution to an
equation of the form
\be
(X_0^2 \ka_0+12 \ka_1+6\ka_2) \nabla^2 A -M^2 X_0^2 \ka_0 A = ... .
\ee
The Green's function for this equation can also be seen to be the same as
the corresponding $O(2)$ case.
The long-ranged components of the massive modes of $h_{00}$ to $O(4)$
are given below.

\bigskip
\noindent
{\bf ISS theory}
\vspace{10pt}
\be
h^{(m)}_{00}
\dot{=} \frac{\pi^2 l^4 G^2}{18
  (1+\sigma)^4} (m+n\sigma )^2(m-n)^2 \rho^2.
\ee

\bigskip
\noindent
{\bf $\mathbf{\ka_0=0}$ theory}
\vspace{10pt}

In this case the massive modes $h^{(m)}_{00}$ and $h^{(m)}_{ij}$ to
$O(2)$ both have long-ranged components, and so the full expressions
do not simplify greatly when the exponentially suppressed modes are
neglected.  The reader is therefore referred to
equations (\ref{hm42}) and (\ref{A42}) in Appendix B, where
$h^{(m)}_{00}$ to $O(4)$ is given explicitly.

\bigskip
\noindent
{\bf Other theories}
\vspace{10pt}
\be
h^{(m)}_{00}\dot{=} 0.
\ee

\bigskip
\noindent
To find these expressions we have made use of (\ref{id1}) and
(\ref{id2}), as well as the following relations:
\bea
\mathcal{W}_M(\mathcal{W}_N(X)) = \frac{4 \pi}{(M^2-N^2)}
\left( \mathcal{W}_N(X)-\mathcal{W}_M(X) \right)
\\ \mathcal{W}_M (\mathcal{V}(X)) \dot{=} \frac{4 \pi}{M^2}
\mathcal{V}(X).
\eea

\section{Perturbations Coupled to Matter Fields}
\label{coupled}

We have so far calculated the form of the massless and massive
combinations of the perturbations to $g_{\mu\nu}$ and $q_{\mu\nu}$.
We now want to know what these tell us about the perturbations to the metric
$\hat{g}_{\mu\nu}$.  Using previous relations we
can write $\hat{g}_{\mu\nu}$ as
\be
\hat{g}_{\mu\nu} = \hat{\eta}_{\mu\nu} + \hat{h}_{\mu\nu}
\ee
where
\be
\hat{h}_{\mu\nu} = \frac{(m+nX_0^2)}{(1+X_0^2)} h^{(0)}_{\mu\nu}
+\frac{X_0^2 (m-n)}{(1+X_0^2)} h^{(m)}_{\mu\nu}
\ee
and $\hat{\eta}_{\mu\nu}=\eta_{\mu\nu}+\bar{\eta}_{\mu\nu}$.  It is
now straightforward to write down the components of
$\hat{h}_{\mu\nu}$.
To proceed in doing this it is useful to define two new constants:
\bea
G_N &\equiv& \frac{G (m+nX_0^2)^3}{(1+X_0^2)}\\
\mathcal{G}_N &\equiv& \frac{G (m+n X_0^2)}{3(1+X_0^2)} \left[ 3 (m+n
  X_0^2)^2 -X_0^2 (m-n)^2 \right].
\eea
The reason for doing this is that although the constant $G$ appears in
the gravitational action (\ref{general}) in the same way as it does in general relativity, it
is {\it not} necessarily the value of Newton's constant that is
determined by Cavendish type experiments.  For the $\ka_0=0$
theories this value of Newton's constant is given by $\mathcal{G}_N$
(when $M$ is large), while for the ISS theory and all other
theories it is given by $G_N$ (when
$M$ is large for the ISS theory, and when both $M$ and $N$ are
large for all other theories).

It is also useful at this point to start using infinitesimal gauge
transformations to put our results in the `standard post-Newtonian gauge'.
In the preceding section we used the gauge specified by
conditions (\ref{gauge2}) and (\ref{gauge3}).  This has been a
convenient choice, and has allowed integration of the field equations
to post-Newtonian accuracy.  It has, however, resulted in the metrics
in question having off-diagonal components in the spatial part of the
metric, as well as non-zero $O(1)$ terms in the $0$-$i$ components of
the metric.  These can be removed by making an infinitesimal
coordinate transformation of the form $x^{\mu} \rightarrow
x^{\mu} + \xi^{\mu}$.  The metric that couples to matter,
$\hat{g}_{\mu \nu}$, is then transformed in such a
way that\footnote{This is true as long the matter does not couple to
  the massive combination of metrics only, i.e. as long as $m\neq -nX_0^2$.}
\be
\hat{h}_{\mu \nu} \rightarrow \hat{h}_{\mu \nu} - \xi_{\mu ; \nu} - \xi_{\nu ;
  \mu} + O(\xi^2),
\ee
where covariant derivatives are with respect to $\hat{g}_{\mu\nu}$,
and where indices have been lowered with $\hat{g}_{\mu\nu}$.  By making
coordinate transformations such that $\xi_0 \sim O(1)$ or $O(3)$ and $\xi_i \sim O(2)$
the metric perturbations then transform as
\bea
\hat{h}_{ij} &\rightarrow& \hat{h}_{ij} -2 \xi_{(i,j)}
\\
\hat{h}_{0i} &\rightarrow& \hat{h}_{0i} - \xi_{0,i} - \xi_{i,0}.
\eea
Transformations of this kind will be used in what follows to remove all
contributions from the $O(1)$ massive modes in $\hat{h}_{0i}$, as well as to diagonalise $\hat{h}_{ij}$, and to remove the
potentials found above that contain derivatives on the functions in
their integrands.

As in the previous section, many of the expressions involved here are
quite lengthy.  We will therefore once again present the full equation
in Appendix C, and quote here only the terms that are long-ranged
when $M$ and $N$ are large (after the gauge transformations discussed
above have been performed\footnote{These transformations are given
  explicitly in Appendix C.}).
To $O(4)$ in $\hat{h}_{00}$,  $O(3)$ in $\hat{h}_{0i}$ and $O(2)$ in
$\hat{h}_{ij}$ we find the results below.
\bigskip


\noindent
{\bf ISS theory}
\vspace{10pt}
\bea
\hspace{-30pt}
\hat{h}_{00} \dot{=} 2 G_N \left[U
  -\frac{G_N}{(m+n\sigma)} U^2+2
  \Phi_1+ \frac{2 G_N}{(m+n\sigma)}  \Phi_2 +\Phi_3 +3 \Phi_4
  \right]
\nonumber \\ \label{h00pf}
+\frac{4 \pi \sigma G^2_N l^2 (m-n)^2}{3
  (1+\sigma)(m+n\sigma)^3}\mathcal{V}(\rho^2)
+ \frac{\pi^2 l^4 G^2_N \sigma (m-n)^3}{18
  (1+\sigma)^3 (m+n\sigma)} \rho^2\\\label{h0ipf}
\hspace{-30pt} \hat{h}_{0i}
\dot{=} -\frac{7G_N}{2} V_i - \frac{G_N}{2} W_i
\\ \hspace{-30pt} \hat{h}_{ij}
\dot{=} 2 G_N \delta_{ij} U.\label{hijpf}
\eea

\newpage

\vspace{10pt}
\noindent
{\bf $\mathbf{\ka_0=0}$ theory}
\vspace{10pt}
\bea
\hspace{-30pt}
\hat{h}_{00}
\dot{=} 2\mathcal{G}_N U +O(4) \label{h00k}
\\\hspace{-30pt} \hat{h}_{0i}
\dot{=} -\frac{7\mathcal{G}_N}{2} \frac{3(m+n X_0^2)^2}{
  (3(m+nX_0^2)^2-X_0^2(m-n)^2)} V_i \nonumber \\  - \frac{\mathcal{G}_N}{2} \frac{3
  (m+n X_0^2)^2}{(3(m+nX_0^2)^2-X_0^2(m-n)^2)} W_i \label{h0ik}
\\\hspace{-30pt} \hat{h}_{ij}
\dot{=} 2 \mathcal{G}_N \frac{\left( 3(m+n
X_0^2)^2  +X_0^2 (m-n)^2  \right)}{\left( 3(m+n
X_0^2)^2  -X_0^2 (m-n)^2  \right)} \delta_{ij} U.\label{hijk}
\eea

\vspace{10pt}
\noindent
{\bf Other theories}
\vspace{10pt}
\bea
\hspace{-30pt} \label{h00g}
\hat{h}_{00} \dot{=} 2 G_N \left[U
  -\frac{G_N}{(m+nX_0^2)} U^2+2
  \Phi_1+ \frac{2 G_N}{(m+nX_0^2)}  \Phi_2 +\Phi_3 +3 \Phi_4
  \right]\\
\hspace{-30pt} \label{h0ig}
\hat{h}_{0i}
\dot{=} -\frac{7G_N}{2} V_i - \frac{G_N}{2} W_i
\\ \hspace{-30pt}\hat{h}_{ij} \dot{=} 2 G_N \delta_{ij} U. \label{hijg}
\eea
\vspace{-5pt}

\noindent
In the $\ka_0=0$ theory we have not written the $O(4)$ part of
$\hat{h}_{00}$ explicitly, because, as mentioned before, it contains a large number of
long-ranged potentials.  Appendices B and C contain enough information
for the reader to calculate these terms explicitly, if they are required.

\section{Post-Newtonian Parameters, and Observational Constraints}
\label{obs}

It is convenient at this point to choose units so that $m+n X_0^2=1$.
The unperturbed metric $\hat{\eta}_{\mu\nu}$ then takes its usual
form.  Using the results of the previous section we can now obtain the
PPN parameters of the theories we are considering, and determine the
extent to which they can be observationally constrained.

\bigskip
\noindent
{\bf ISS theory}
\vspace{10pt}

It can be seen by comparing (\ref{h00pf}), (\ref{h0ipf}) and
(\ref{hijpf}) with (\ref{pnm1}), (\ref{pnm2}) and (\ref{pnm3}) that,
when $M$ is large, all PPN parameters take the same value as in General Relativity. We
do, however, have two new potentials in the $\hat{h}_{00}$ term at
$O(4)$.  These are
\be
\label{pfanom}
\hspace{-45pt}
+\frac{4 \pi \sigma G^2_N l^2 (1-n(1+\sigma))^2}{3
  (1+\sigma)}\mathcal{V}(\rho^2)
\qquad \textrm{and} \qquad
+ \frac{\pi^2 l^4 G^2_N \sigma (1-n(1+\sigma))^3}{18
  (1+\sigma)^3} \rho^2.
\ee
No such terms appear in the usual PPN prescription.  The second
of these is particularly unfamiliar as it involves $\rho$ directly, and not
$\rho$ integrated over any volume.  Such a term, however, can be
expected to be small in comparison to $\mathcal{V}(\rho^2)$ so long as $l
\ll d$, where $d$ is the length scale over which the observational
phenomena are being measured.  Outside of any matter distribution, of
course, this term will vanish entirely.

In the absence of any observational constraints on these new
potentials, the only constraint that can currently be placed
on this particular theory is therefore $M d \gg 1$, or equivalently
\be
4 (1+\sigma) \gg \frac{l^2}{d^2}.
\ee
No other constraints are available from standard PPN formalism.

\bigskip
\noindent
{\bf $\mathbf{\ka_0=0}$ theory}
\vspace{10pt}

In the $\ka_0=0$ theory it can be seen by comparing (\ref{h00k}),
(\ref{h0ik}) and (\ref{hijk}) with (\ref{pnm1}), (\ref{pnm2}) and
(\ref{pnm3}) that, among other deviations from general relativity, we have
\be
\gamma = \frac{\left(3 +X_0^2(1-n(1-X_0^2))^2  \right)}{\left(3 -X_0^2 (1-n(1-X_0^2))^2  \right)}.
\ee
As in the ISS theory, the $\ka_0=0$ theory must also obey $M d
\gg 1$ in order for the massive modes to be exponentially suppressed.
For this theory, this corresponds to
\be
\frac{8 (1+X_0^2)}{X_0^2} (2 \ka_1+\ka_2) \gg \frac{l^2}{d^2}.
\ee
Unlike the ISS theory, however, we can also use the PPN
constraint on $\gamma$ from observations of the Cassini spacecraft to
constrain the theory.  To be compatible with observations $\gamma$
must satisfy equation (\ref{refgam}).  We then have the constraint
\be
\label{k0con}
X_0^2(1-n(1-X_0^2))^2 = (3.1 \pm  3.3) \times 10^{-5}.
\ee
While tight, it is worth noting that this constraint can always be
evaded by an appropriate choice of $X_0$.  Unless $X_0$ is known from
other considerations (cosmology, for example) it is not possible to
constrain the theory with this information only.  It is also worth
noting that this is a constraint on $n$ only, and not the parameters
determining the interactions in the theory (i.e. $\ka_1$ and $\ka_2$).
Further constraints are also available on this theory from the $\hat{g}_{00}$
metric component at $O(4)$.


\bigskip
\noindent
{\bf Other theories}
\vspace{10pt}

Comparing (\ref{h00g}), (\ref{h0ig}) and (\ref{hijg}) with
(\ref{pnm1}), (\ref{pnm2}) and (\ref{pnm3}) shows, as long as
both $M d \gg 1$ and $N d\gg 1$, that all of the PPN parameters in
this general class of theories are the same as in general relativity.
The only constraints that need to be satisfied by these theories are therefore
\be
\frac{2 (1+X_0^2)}{X_0^2} (X_0^2 \ka_0+8\ka_1+4\ka_2) \gg \frac{l^2}{d^2}
\ee
and
\be
2 \ka_0 (1+X_0^2) \frac{(X_0^2 \ka_0+8\ka_1+4\ka_2)}{(X_0^2
\ka_0+12\ka_1+6\ka_2)} \gg \frac{l^2}{d^2} .
\ee
If these conditions are met then all the post-Newtonian predictions of
these theories are indistinguishable from those of general
relativity.  This shows the importance of the $\ka_0$ term in the
interaction Lagrangian.

\section{Discussion}
\label{disc}

We have investigated here the weak field limit of a general class of
bimetric theories, generated from the gravitational action
(\ref{general}), and coupled to matter via a linear combination
of the two metrics involved.  We have calculated the post-Newtonian
limit of these theories, up to $O(4)$ in $g_{00}$, $O(3)$ in $g_{0i}$
and $O(2)$ in $g_{ij}$, for a perfect fluid matter content distributed
without any symmetries.  We find that, to all orders considered,
there is a natural decomposition of the perturbations to the two
metrics into modes we have called `massive' and `massless'.  These
modes are defined through the combinations
\be
h^{(m)}_{\mu \nu} = h_{\mu \nu}-\frac{\hb_{\mu \nu}}{X_0^2} \qquad
\mathrm{and} \qquad h^{(0)}_{\mu \nu} = h_{\mu \nu} +\hb_{\mu \nu},
\ee
where $h_{\mu \nu}$ and $\hb_{\mu \nu}$ denote the perturbations to
the two metrics involved, and $X_0$ is a constant associated with the
unperturbed background space-times.

We find that, in general, the $O(1)$ contributions to the massive
modes $h^{(m)}_{0i}$ are non-zero, and cannot be made to vanish by a
coordinate transformation.  These quantities satisfy a Helmholtz
equation with mass $M$ (as given by Eq. (\ref{M})), and while the $0$-$i$
component of the combined metric that couples to matter can always be
set to zero by a gauge transformation (as long as the matter does not
couple exclusively to the massive modes), these $O(1)$ terms can in principle make a
non-negligible contribution to the other components of the metric.
We therefore include the effects of these terms while calculating the
$0$-$0$ component of the combined metric to $O(2)$, but
neglect them to higher orders, as we see no reason to expect them to
contribute substantially to post-Newtonian accuracy if they are
negligible at the Newtonian level.

To Newtonian order, we find that in general the relevant metric perturbations
have massless modes (that are always long-ranged), as well as massive
modes with {\it two} different masses: $M$ and $N$ (where $N$ is given
by Eq. (\ref{N})).  We find two special cases in which the second
mass, $N$, does not appear.  These are the ISS theory, as
specified in Section \ref{bi}, and theories with $\ka_0=0$, in which the interaction
term $\sqrt{-q} \qin^{\mu \nu} g_{\mu \nu}$ is missing from the
action.  In the ISS theory this occurs because the $\nabla h^{(m)}_{00}=...$
equation can be decoupled from the $h^{(m)}_{ij}$ terms without having
to solve any other differential equations.  In all other cases this is
not possible, and a second equation of the form $\nabla^2 A -N^2
A=...$ must be solved, before one can integrate the $h^{(m)}_{00}$
equation.  In the second anomalous case, when $\ka_0\rightarrow 0$ it is found that the mass
parameter $N\rightarrow 0$.  Theories of this kind therefore have extra massless
modes, beyond the usual Newtonian term that occurs from integrating
the $\nabla h^{(0)}_{00}=...$ equation.

We then proceed to calculate the metric perturbations to
post-Newtonian accuracy.  These terms all involve the same two masses,
$M$ and $N$, that are required to Newtonian order (except for the
ISS and $\ka_0=0$ cases, which only require $M$).  By
expressing these perturbations in the standard post-Newtonian gauge we find
that the predictions of these theories are, in general, very similar
to those of general relativity.  In particular, if $M$ and $N$ are
both large enough then they can evade all preferred frame experiments,
and are indistinguishable from general relativity by any experiments
that constrain the PPN parameters $\beta$ and $\gamma$ (including the
stringent tests available from Shapiro time-delay observations \cite{gamma},
and lunar laser ranging measurements \cite{beta}).  The ISS
theory also does a remarkably good job of satisfying weak field tests
of gravity when $M$ is large, with PPN parameters that are again the same as in general
relativity.  In this case, however, we do find a couple of anomalous
terms appearing at $O(4)$ in the $0$-$0$ component of the metric that
couples to matter.  These are given in Eq. (\ref{pfanom}), and could
potentially be used to observationally distinguish between general
relativity and ISS theory.  The
$\ka_0=0$ theory does not fare quite so well, due to its extra long-ranged
modes.  In this case observations of $\gamma$ can be used to place
constraints on a particular combination of the constant $X_0$, and the
mixing angle between the two fundamental spin-2 fields that
couples to matter, even when $M$ is large.

As well as the large mass limit, one can also consider the small mass
limit.  In this case a viable Newtonian term can again appear in the $0$-$0$
component of the metric, and we must then consider the problem of the vDVZ
discontinuity that was discussed in Section \ref{intro}.  To do this
we note that as $M\rightarrow 0$ and $N\rightarrow 0$ we get
$\mathcal{W}_M(\rho) \rightarrow U^*$ and $\mathcal{W}_N(\rho)
\rightarrow U^{\dagger}$, respectively (we have used the superscripts
$*$ and $\dagger$ here, even though $U^*=U^{\dagger}=U$, so that we
can keep track of where the potential has come from).  Neglecting
contributions from $h^{(m)}_{0i}$ at $O(1)$, the full expressions in
Appendix C then give, for the most general theories,
\bea
\hspace{-35pt}
\hat{h}_{00} = 2 G_N U + \frac{8 G_N X_0^2 (1-n (1+X_0^2))^2 U^*}{3} -\frac{2
  G_N X_0^2 (1-n (1+X_0^2))^2 U^{\dagger}}{3} \\ \hspace{-35pt}
\hat{h}_{ij} = 2 G_N U + \frac{4 G_N X_0^2 (1-n (1+X_0^2))^2 U^*}{3}
+\frac{2 G_N X_0^2 (1-n (1+X_0^2))^2 U^{\dagger}}{3}.
\eea
The vDVZ discontinuity can be seen to be manifest in the $U^*$ terms
of these two equations.  If these were the only terms in these
expressions then we would indeed get the familiar $\gamma =1/2$ that
is anticipated from massive gravity.  Here, however, we have two
additional terms in both $\hat{h}_{00}$ and $\hat{h}_{ij}$ -- one from
the terms that are long-ranged even when $M$ and $N$ are large, and
another from the $\mathcal{W}_N(\rho)$ terms.  The cumulative effect
of all three of these contributions added together shows that there is
no vDVZ discontinuity in the $\gamma$ parameter of the most general class
of these theories, as in this case we end up with $\hat{h}_{ij} =
\delta_{ij} \hat{h}_{00}$, and hence $\gamma =1$.  The same is true in
the special case in which $\ka_0=0$, as should be expected, as these
modes already correspond to the $N \rightarrow 0$ limit of the more
general theories.  The ISS theory, however, does not have $\gamma
= 1$ as $M \rightarrow 0$.  Instead we find
\be
\gamma \rightarrow \frac{3+2 (1-n (1+\sigma))^2\sigma}{3+4 (1-n (1+\sigma))^2\sigma}.
\ee
This is essentially because the
$\mathcal{W}_N(\rho)$ terms that cancel the discontinuity from the
$\mathcal{W}_M(\rho)$ terms are absent in the ISS theory.

It should be noted, however, that although the general relativistic
value of $\gamma$ is recovered in the most general theories as $M$ and
$N \rightarrow 0$, this does not mean that all PPN parameters will
be.  For example, for the parameters $\alpha_1$ and $\alpha_2$ that
parameterise deviations from general relativity in the $0$-$i$
components of the field equations, we find that while $\alpha_1
\rightarrow 0$ (as in general relativity), we also have
\be
\label{alpha2} \alpha_2 \rightarrow -\frac{(1-n (1+X_0^2))^2
  X_0^2}{(2-n (1+X_0^2))^2 X_0^2}.
\ee
In terms of this expression, equation
(\ref{refalph}) then gives the constraint
\be
X_0^2 (1-n (1+X_0^2))^2 \lesssim 1.2 \times 10^{-7}.
\ee
This is the same combination of quantities that was
constrained in the large $M$ limit of the $\ka_0=0$ theory using the
parameter $\gamma$, but the constraint here is from an entirely
different gravitational effect, and is even tighter.  On should also
note that in the limit $M \rightarrow 0$ the $O(1)$ terms in
$h^{(m)}_{0i}$ are no longer short-ranged.  These terms are, however,
set by initial conditions, rather than by the matter content of the
space-time.  One may then be able to limit their magnitude by applying
suitable observational constraints, but this will place constraints on
the initial conditions only, and not on the gravitational theory itself.

Finally, we note that when the matter fields are coupled to the
massless combination of metrics only (so that $m=n$) then many of the
constraints found above are automatically satisfied.  These
observations can therefore be though of as constraining the amount of
massive mode that is allowed to be mixed with the massless modes.



\section*{Acknowledgements}

TC acknowledges the support of Jesus College, Oxford and the BIPAC.
MB would like to thank the Oxford Astrophysics department for hospitality
during a sabbatical year. MB was partially supported by the
JS Guggenheim Memorial Foundation,  Fondecyt (Chile) Grant \# 1100282 and
Alma-Conicyt Grant \# 31080001.

\newpage
\section*{Appendix A:  Perturbed Metric Quantities}

The perturbed form of the metric that we will be using is specified in
(\ref{gandq1}) and (\ref{gandq2}).   In order to solve
(\ref{R}) and (\ref{K}), however, we also need to know the perturbed Ricci tensor
to $O(3)$ in its off-diagonal components, and $O(4)$ in all of its diagonal
components\footnote{The $i$-$j$ components are usually only required to
  $O(2)$, but here we also need higher order terms in order to
  calculate the $O(4)$ part of the $0$-$0$ components.}, as well as the
relevant expressions for the perturbed metric determinants, and
inverse metrics.
The perturbed Ricci tensor components take on their usual lengthy
form, and we will not reproduce them explicitly here.

The perturbed metric determinants can be calculated
using the equation $det(g)=e^{tr(ln(g))}$. To $O(2)$ in perturbations this gives
\bea
det(g)
&=& -\left(1- h_{00}+h_{ii}+h_{0i}^2 \right) +O(4) \\
det(q) &=& -X_0^8 \left(1- \frac{\hb_{00}}{X_0^2}+
\frac{\hb_{ii}}{X_0^2} +\frac{\hb_{0i}^2}{X_0^4} \right) +O(4)
\\
det(\hat{g}) &=& -  \Bigg[ (m+n X_0^2)^2 +
  (h_{ii}-h_{00}+h_{0i}^2) m^2 \nonumber \\&& \qquad
  +(X_0^2 h_{ii}+\hb_{ii}-X_0^2 h_{00}-\hb_{00} +2
  h_{0i}\hb_{0i}) mn \nonumber \\&& \qquad +(X_0^2 \hb_{ii}-X_0^2
  \hb_{00}+\hb_{0i}^2) n^2  \Bigg](m+n X_0^2)^2 +O(4),
\eea
and, if we ignoring $O(1)$ contributions to $h_{0i}$ and $\hb_{0i}$,
then to higher order we have
\bea
det(g) &=& -\left(1 -h_{00} +h_{ii} -h_{00} h_{ii} +\frac{1}{2}
  h_{ii}^2 -\frac{1}{2} h_{ij}^2\right) +O(6)\\
det(q) &=& -X_0^8 \Bigg( 1 - \frac{\hb_{00}}{X_0^2}
+\frac{\hb_{ii}}{X_0^2}
  -\frac{\hb_{00} \hb_{ii}}{X_0^4} + \frac{\hb_{ii}^2}{2 X_0^4}
  -\frac{\hb_{ij}^2}{2 X_0^4} \Bigg) +O(6).
\eea

In order to compute the interaction terms in the field equations
(\ref{R}) and (\ref{K}) we also need to know the inverse metric
fluctuations $\qin^{\mu \nu}= \etain^{\mu \nu} + \dqin^{\mu \nu}$.  These can be found from the definition
$\qin^{\mu \alpha} q_{\alpha \nu} = \delta^{\mu}_{\nu}$, which gives
\be
\hspace{-20pt}
\dqin^{\mu \nu} = -\etain^{\mu \alpha} \etain^{\nu \beta} \hb_{\alpha
  \beta} + \etain^{\mu \alpha} \etain^{\nu \delta}\etain^{\beta \gamma} \hb_{\alpha \beta}\hb_{\gamma \delta} +O(h^3).
\ee
The components of $\dqin^{\mu \nu}$ are then given by
\bea
\dqin^{00} &=& -\frac{\hb_{00}}{X_0^4} + \frac{\hb_{0i}^2}{X_0^6} +O(4)\\
\dqin^{0i} &=& \frac{\hb_{0i}}{X_0^4}+O(3)\\
\dqin^{ij} &=& -\frac{\hb_{ij}}{X^4_0} - \frac{\hb_{0i} \hb_{0j}}{X_0^6}+O(4).
\eea
Ignoring $O(1)$ contributions to $h_{0i}$ and $\hb_{0i}$, 
the higher order components of $\dqin^{\mu \nu}$ are 
\bea
\dqin^{00} &=& -\frac{\hb_{00}}{X_0^4} -\frac{\hb_{00}^2}{X_0^6} +O(6)\\
\dqin^{ij} &=& -\frac{\hb_{ij}}{X^4_0}
+\frac{\hb_{ik}\hb_{jk}}{X_0^6} +O(6).
\eea

\newpage
\section*{Appendix B: The {\boldmath $g^{(m)}_{00}$} and {\boldmath
    $q^{(m)}_{00}$} Equations, to {\boldmath $O(4)$}}

Using this expression, together with other known identities and lower
  order solutions, the field equations (\ref{R}) and (\ref{K}) give
\bea
 \nabla^2 h^{(m)}_{00} -M^2 h^{(m)}_{00}
\nonumber \\ \hspace{-15pt} =
\frac{2 (1+X_0^2)(2\ka_1+\ka_2)}{l^2X_0^2} A - \frac{4 (1+X_0^2)
  (2\ka_1+\ka_2)}{M^2l^2 X_0^2} A_{,00}
+ h^{(m)}_{00,00}
\nonumber \\
-8\pi G (m+nX_0^2) (m-n)\rho \left[1+\Pi+2
  v^2 +3\frac{p}{\rho}\right] -\frac{2(2 \ka_1+\ka_2)}{M^2l^2X_0^2} h^{(0)}_{00,i} A_{,i}
\nonumber \\
+ \frac{8 \pi G (m+nX_0^2) (mX_0^2+n)}{(1+X_0^2)}\rho
  h^{(m)}_{00}
+ 4 \pi G  m n(1+X_0^2) \rho \left( h^{(m)}_{00} + h^{(m)}_{ii} \right)
 \nonumber \\
+\frac{1}{(1+X_0^2)} \left[h^{(0)}_{00,ij} h^{(m)}_{ij} -2
  h^{(0)}_{00,i} h^{(m)}_{00,i}  \right] + \frac{2 (X_0^2 \ka_0+4
  \ka_1+2\ka_2)}{l^2 X_0^2} h^{(0)}_{00}h^{(m)}_{00}
 \nonumber \\
+\frac{(1-X_0^2)}{2 (1+X_0^2)} \left[ h^{(m)2}_{00,i} -
  2h^{(m)}_{00,ij} h^{(m)}_{ij} - 2 h^{(m)}_{00,i} h^{(m)}_{ij,j}
  +h^{(m)}_{00,i} h^{(m)}_{jj,i} \right]+\frac{\ka_1}{l^2 X_0^2}
h^{(m)^2}_{ii}
 \nonumber \\
-\frac{\left[X_0^2 \ka_0+8 \ka_1 +6 \ka_2\right]}{4 l^2}  \left(
3 h^{(m)2}_{00} +2 h^{(m)}_{00} h^{(m)}_{ii} + h^{(m)2}_{ii}-2 h^{(m)2}_{ij}\right)
 \nonumber \\
-\frac{3 (\ka_1+\ka_2)}{l^2 X_0^2} h^{(m)2}_{00}
+\frac{2 (3\ka_1+\ka_2)}{l^2 X_0^2}  h^{(m)}_{00} h^{(m)}_{ii}
+\frac{(4\ka_1+3\ka_2)}{l^2 X_0^2} h^{(m)2}_{ij}.
\label{hm004}
\eea
The solutions to this equation are given below.

\vspace{15pt}
\noindent
{\bf ISS theory}
\bea
\hspace{-40pt} h^{(m)}_{00}
=
\frac{(1+\sigma)}{2 \pi l^2} \mathcal{W}_M(A) -\frac{\mathcal{W}_M(A_{,00})}{4\pi}
- \frac{2 G (m+n\sigma) (m\sigma+n)}{(1+\sigma)}\mathcal{W}_M(\rho
  h^{(m)}_{00})
\nonumber \\
-G  m n(1+\sigma)  \left[ \mathcal{W}_M(\rho h^{(m)}_{00}) +
\mathcal{W}_M(\rho h^{(m)}_{ii} )\right]
-\frac{ \mathcal{W}_M(h^{(0)}_{00,i}A_{,i})}{8\pi(1+\sigma )}
\nonumber \\
+2 G (m+n\sigma) (m-n) \left[\mathcal{W}_M(\rho)+\mathcal{W}_M(\rho \Pi)+2
  \mathcal{W}_M(\rho v^2) +3 \mathcal{W}_M(p)\right]
 \nonumber \\
-\frac{( \mathcal{W}_M(h^{(0)}_{00,ij} h^{(m)}_{ij}) -2
   \mathcal{W}_M(h^{(0)}_{00,i} h^{(m)}_{00,i}))}{4 \pi (1+\sigma)}
-\frac{ \mathcal{W}_M(h^{(m)}_{00,00})}{4\pi}
 \nonumber \\
-\frac{\left[ 8  \mathcal{W}_M(h^{(0)}_{00}h^{(m)}_{00})
-4   \mathcal{W}_M(h^{(m)}_{00} h^{(m)}_{ii} )
- \mathcal{W}_M(h^{(m)^2}_{ii})
- \mathcal{W}_M(h^{(m)2}_{ij}) \right]}{4\pi l^2}
 \nonumber \\
+\frac{\sigma}{4 \pi l^2}  \left[
3 \mathcal{W}_M(h^{(m)2}_{00}) +2 \mathcal{W}_M(h^{(m)}_{00}
h^{(m)}_{ii}) + \mathcal{W}_M(h^{(m)2}_{ii})-2
\mathcal{W}_M(h^{(m)2}_{ij})\right]
 \nonumber \\
-\frac{(1-\sigma)}{8\pi (1+\sigma)} \Big[ \mathcal{W}_M(h^{(m)2}_{00,i}) -
  2\mathcal{W}_M(h^{(m)}_{00,ij} h^{(m)}_{ij}) \nonumber \\ \qquad
  \qquad \qquad \qquad- 2 \mathcal{W}_M(h^{(m)}_{00,i} h^{(m)}_{ij,j})
  +\mathcal{W}_M(h^{(m)}_{00,i} h^{(m)}_{jj,i}) \Big] .
\label{hm41}
\eea

\newpage
\noindent
{\bf $\mathbf{\ka_0=0}$ theory}
\bea
\hspace{-40pt}h^{(m)}_{00} =
-\frac{(1+X_0^2)(2\ka_1+\ka_2)}{2 \pi l^2X_0^2} \mathcal{W}_M(A) + \frac{\mathcal{W}_M(A_{,00})}{8\pi}
+\frac{\mathcal{W}_M(h^{(0)}_{00,i} A_{,i})}{16\pi (1+X_0^2)}
\nonumber \\
+2 G (m+nX_0^2) (m-n) \left[\mathcal{W}_M(\rho)+\mathcal{W}_M(\rho \Pi)+2
  \mathcal{W}_M( \rho v^2) +3 \mathcal{W}_M(p)   \right]
\nonumber \\
- \frac{2 (m+nX_0^2) (mX_0^2+n)}{(1+X_0^2)}\mathcal{W}_M(\rho
  h^{(m)}_{00})
-\frac{\mathcal{W}_M( h^{(m)}_{00,00})}{4\pi}
\nonumber \\
- G  m n(1+X_0^2) \left[ \mathcal{W}_M(\rho h^{(m)}_{00}) +\mathcal{W}_M(\rho h^{(m)}_{ii}) \right]
 \nonumber \\
-\frac{(\mathcal{W}_M(h^{(0)}_{00,ij} h^{(m)}_{ij}) -2
  \mathcal{W}_M(h^{(0)}_{00,i} h^{(m)}_{00,i})  )}{4 \pi (1+X_0^2)}  - \frac{(2
  \ka_1+\ka_2)}{\pi l^2 X_0^2} \mathcal{W}_M(h^{(0)}_{00}h^{(m)}_{00})
 \nonumber \\
+\frac{(4 \ka_1 +3 \ka_2)}{8\pi l^2}  \left[
3 \mathcal{W}_M(h^{(m)2}_{00}) +2 \mathcal{W}_M(h^{(m)}_{00}
h^{(m)}_{ii}) + \mathcal{W}_M(h^{(m)2}_{ii})-2
\mathcal{W}_M(h^{(m)2}_{ij}) \right]
 \nonumber \\
+\frac{3 (\ka_1+\ka_2)}{4\pi l^2 X_0^2} \mathcal{W}_M(h^{(m)2}_{00} )
-\frac{(3\ka_1+\ka_2)}{2 \pi l^2 X_0^2} \mathcal{W}_M( h^{(m)}_{00}
h^{(m)}_{ii} )
 \nonumber \\
-\frac{\ka_1}{4 \pi l^2 X_0^2}\mathcal{W}_M(h^{(m)^2}_{ii})
-\frac{(4\ka_1+3\ka_2)}{4 \pi l^2 X_0^2} \mathcal{W}_M(h^{(m)2}_{ij})
 \nonumber \\
-\frac{(1-X_0^2)}{8\pi (1+X_0^2)} \Big[ \mathcal{W}_M(h^{(m)2}_{00,i}) -
  2\mathcal{W}_M(h^{(m)}_{00,ij} h^{(m)}_{ij})
\nonumber \\
\qquad \qquad \qquad \qquad - 2\mathcal{W}_M( h^{(m)}_{00,i} h^{(m)}_{ij,j})
  +\mathcal{W}_M(h^{(m)}_{00,i} h^{(m)}_{jj,i}) \Big].
\label{hm42}
\eea

\noindent
{\bf Other theories}
\bea
\hspace{-40pt}h^{(m)}_{00} =
-\frac{(1+X_0^2)(2\ka_1+\ka_2)}{2 \pi l^2X_0^2} \mathcal{W}_M(A) + \frac{(1+X_0^2)
  (2\ka_1+\ka_2)}{\pi M^2l^2 X_0^2} \mathcal{W}_M(A_{,00} )
\nonumber \\
-\frac{\mathcal{W}_M( h^{(m)}_{00,00})}{4\pi} +\frac{(2
  \ka_1+\ka_2)}{2\pi M^2l^2X_0^2} \mathcal{W}_M(h^{(0)}_{00,i} A_{,i})
\nonumber \\
+2 G (m+nX_0^2) (m-n) \left[\mathcal{W}_M(\rho)+\mathcal{W}_M(\rho \Pi)+2
  \mathcal{W}_M(\rho v^2) +3 \mathcal{W}_M(p)  \right]
\nonumber \\
               - \frac{2 G (m+nX_0^2) (mX_0^2+n)}{(1+X_0^2)}\mathcal{W}_M(\rho
  h^{(m)}_{00})-\frac{(4\ka_1+3\ka_2)}{4\pi l^2 X_0^2} \mathcal{W}_M(h^{(m)2}_{ij})
\nonumber \\
- G  m n(1+X_0^2)  \left[ \mathcal{W}_M(\rho h^{(m)}_{00})
  +\mathcal{W}_M(\rho h^{(m)}_{ii}) \right]
-\frac{\ka_1}{4\pi l^2 X_0^2}\mathcal{W}_M( h^{(m)^2}_{ii})
 \nonumber \\
-\frac{(\mathcal{W}_M(h^{(0)}_{00,ij} h^{(m)}_{ij}) -2
  \mathcal{W}_M(h^{(0)}_{00,i} h^{(m)}_{00,i} ) )}{4 \pi (1+X_0^2)}
-\frac{(3\ka_1+\ka_2)}{2 \pi l^2 X_0^2} \mathcal{W}_M( h^{(m)}_{00}
h^{(m)}_{ii} )
\nonumber \\
- \frac{(X_0^2 \ka_0+4
  \ka_1+2\ka_2)}{2 \pi l^2 X_0^2}
\mathcal{W}_M(h^{(0)}_{00}h^{(m)}_{00})
+\frac{3 (\ka_1+\ka_2)}{4\pi l^2 X_0^2} \mathcal{W}_M(h^{(m)2}_{00})
 \nonumber \\
-\frac{(1-X_0^2)}{8\pi (1+X_0^2)} \Big[ \mathcal{W}_M(h^{(m)2}_{00,i}) -
  2\mathcal{W}_M(h^{(m)}_{00,ij} h^{(m)}_{ij}) \nonumber \\ \hspace{5cm}- 2\mathcal{W}_M( h^{(m)}_{00,i} h^{(m)}_{ij,j})
  +\mathcal{W}_M(h^{(m)}_{00,i} h^{(m)}_{jj,i} )\Big]
 \nonumber \\
+\frac{\left[X_0^2 \ka_0+8 \ka_1 +6 \ka_2\right]}{16\pi l^2}  \Big[
3 \mathcal{W}_M(h^{(m)2}_{00}) +2 \mathcal{W}_M(h^{(m)}_{00}
h^{(m)}_{ii})\nonumber \\ \hspace{5cm} +\mathcal{W}_M( h^{(m)2}_{ii})-2
\mathcal{W}_M(h^{(m)2}_{ij}) \Big].
\label{hm43}
\eea
The expressions above should be understood to have
$h^{(0)}_{00}$, $h^{(m)}_{00}$, $h^{(m)}_{ij}$ and $A$ given to $O(2)$ by the
lower order solutions found in previous sections.
It now remains to find $A$ to $O(4)$.  To do this we use the field
  equations (\ref{R}) and (\ref{K}), as well as known identities and
  lower order solutions, to find
\bea
  (X_0^2 \ka_0+12 \ka_1+6\ka_2) \nabla^2 A -M^2
X_0^2 \ka_0 A
\nonumber\\ \hspace{-15pt}=
-\frac{3 M^2 l^2 X_0^2}{2 (1+X_0^2)} h^{(m)}_{00,00}
-\frac{4 \pi G M^2 l^2 X_0^2}{(1+X_0^2)} \left[ 3 m n (1-X_0^2) +\frac{(m-n)^2 X_0^2}{(1+X_0^2)} \right]\rho h^{(m)}_{00}
\nonumber\\
+6 (2\ka_1+\ka_2) A_{,00}
+\frac{4 \pi G M^2 l^2 X_0^2}{3 (1+X_0^2)} \left[  2 m n
  (1+X_0^2) -\frac{(m-n)^2 X_0^2}{(1+X_0^2)}\right] \rho h^{(m)}_{ii}
\nonumber\\
-\frac{8 \pi G M^2 l^2 X_0^2
(m+nX_0^2) (m-n)}{(1+X_0^2)} \rho \left[1+\Pi- 3 \frac{p}{\rho}
  + \frac{2 h^{(0)}_{00}}{(1+X_0^2)}\right]
\nonumber \\
 +\frac{M^2 (2 \ka_1+\ka_2)}{(1+X_0^2)}
\Bigg[
2 h^{(0)}_{00} A +8  h^{(0)}_{00} h^{(m)}_{00}
-4 h^{(m)2}_{00} + A^2 -4 h^{(m)2}_{ij}
\Bigg] \nonumber \\
+\frac{(X_0^2 \ka_0+12\ka_2+6\ka_2)}{(1+X_0^2)} \left[ h^{(0)}_{00}
  \nabla^2 h^{(m)}_{ii}+h^{(m)}_{ii}\nabla^2 h^{(0)}_{00} +2
  h^{(0)}_{00,i} h^{(m)}_{jj,i} \right]
\nonumber \\
-\frac{(X_0^2\ka_0+4\ka_1+2\ka_2)}{(1+X_0^2)} \left[ h^{(0)}_{00}
  \nabla^2 h^{(m)}_{00}+h^{(m)}_{00} \nabla^2 h^{(0)}_{00} \right]
\nonumber \\-\frac{M^2 l^2 X_0^2}{2 (1+X_0^2)^2} \left[ 2 h^{(m)}_{ij}
  h^{(0)}_{00,ij}+5 h^{(0)}_{00,i} h^{(m)}_{ij,j}  \right]+\frac{4 (2\ka_1+\ka_2)}{(1+X_0^2)} h^{(0)}_{00,i} h^{(m)}_{00,i}
\nonumber \\
+\frac{(X_0^2 \ka_0+2 (1+X_0^2) \ka_2)}{(1+X_0^2)}
\left[ h^{(m)}_{ij,k}h^{(m)}_{ij,k} + h^{(m)}_{ij} \nabla^2
h^{(m)}_{ij} \right] \nonumber \\
 +\frac{(X_0^2 \ka_0 +4 (1-X_0^2) \ka_1 +4 \ka_2)}{(1+X_0^2)}
\left[ h^{(m)}_{ij,i} A_{,j}+ h^{(m)}_{ij} A_{,ij}\right]\nonumber \\-
\frac{M^2l^2 X_0^4}{(1+X_0^2)^2} \left[h^{(m)}_{ij,i} h^{(m)}_{00,j}
  +h^{(m)}_{ij} h^{(m)}_{00,ij}\right]+(X_0^2\ka_0 +4\ka_1 +4 \ka_2) \left[ A_{,i}
  h^{(m)}_{ij,j}+A h^{(m)}_{ij,ij}\right]
\nonumber \\
-\frac{(X_0^2 \ka_0+2 \ka_2)}{2} \left[A_{,i}^2
  +A \nabla^2 A-2 h^{(m)2}_{00,i}-2 h^{(m)}_{00}
  \nabla^2 h^{(m)}_{00} \right]
\nonumber \\- \frac{X_0^2 (3 X_0^2 \ka_0+16 \ka_1 +8 \ka_2 )}{(1+X_0^2)}
\left[ h^{(m)2}_{00,i}+h^{(m)}_{00}
  \nabla^2 h^{(m)}_{00} \right]
\nonumber \\
- \frac{2 (X_0^2 \ka_0 +8 \ka_1 +2 (3+X_0^2) \ka_2)}{(1+X_0^2)} \left[
  h^{(m)2}_{ij,j}+  2 h^{(m)}_{ij}
  h^{(m)}_{jk,ki}+h^{(m)}_{ij,k} h^{(m)}_{jk,i}\right]
\nonumber \\
  -\frac{M^2 l^2 X_0^2(1-X_0^2)}{4(1+X_0^2)^2} \Bigg[ 3h^{(m)2}_{00,i} +
  2 h^{(m)}_{00} \nabla^2 h^{(m)}_{00} - A_{,i}^2
  -2 h^{(m)}_{ij,k}h^{(m)}_{ik,j} +2
  h^{(m)}_{jk}\nabla^2h^{(m)}_{jk}
  \nonumber\\  \qquad \qquad \qquad +3
  h^{(m)2}_{ij,k} - 4 h^{(m)2}_{ij,j} + 4 h^{(m)}_{ij,j}
  A_{,i} +2 h^{(m)}_{ij} A_{,ij} -
  4 h^{(m)}_{jk} h^{(m)}_{ij,ik}\Bigg].
\label{A40}
\eea
The solutions to this equation
are given below.

\newpage
\noindent
{\bf ISS theory}
\vspace{15pt}
\bea
\hspace{-70pt} \frac{24 (1+\sigma)}{l^2}  A  =
16 \pi G \left[ 3 m n (1-\sigma ) +\frac{(m-n)^2
    \sigma}{(1+\sigma )}
-\frac{2 (m+n \sigma)^2  }{(1+\sigma )}
+\frac{4 (m+n\sigma) (m-n)}{(1+\sigma )}
\right]\rho h^{(m)}_{00}
\nonumber\\
-\frac{16 \pi G}{3} \left[  2 m n
  (1+\sigma) -\frac{(m-n)^2 \sigma}{(1+\sigma)}
-\frac{3 (3+2 \sigma)(m+n\sigma) (m-n)}{(1+\sigma)}
\right] \rho h^{(m)}_{ii}
\nonumber\\
+ 32 \pi G (m+n \sigma ) (m-n) \rho \left[1+\Pi- 3 \frac{p}{\rho}
  + \frac{h^{(0)}_{00}}{(1+\sigma)}
-\frac{(3+2 \sigma )h^{(m)}_{00} }{6 (1+\sigma )} -\frac{A}{3}
\right]
\nonumber \\
 +\frac{4}{l^2}
\Bigg[12 h^{(0)}_{00} h^{(m)}_{00}
-10 h^{(m)2}_{00}
-2 (5+2 \sigma) h^{(m)2}_{ij}
\Bigg]
-A^2_{,i}
-2 A \nabla^2 A
+6 A_{,00}
+6 h^{(m)}_{00,00}
\nonumber\\
+\frac{2}{(1+\sigma )} \left[ 2 h^{(0)}_{00,i} h^{(m)}_{00,i}    +2 h^{(m)}_{ij}
  h^{(0)}_{00,ij}+5 h^{(0)}_{00,i} A_{,i}  \right]
-5 h^{(m)2}_{ij,k}
+2 h^{(m)}_{ij} A_{,ij}
\nonumber \\
+\frac{4 \sigma}{(1+\sigma)} \left[A_{,i} h^{(m)}_{00,i}
  +h^{(m)}_{ij} h^{(m)}_{00,ij}\right]
-\frac{(5+\sigma)}{(1+\sigma)}  h^{(m)2}_{00,i}
+ 6 h^{(m)}_{ij,k} h^{(m)}_{jk,i}.
\label{A41}
\eea

\noindent
{\bf $\mathbf{\ka_0=0}$ theory}
\vspace{15pt}
\bea
\hspace{-25pt} A =\frac{4 G}{3} \left[ 3 m n (1-X_0^2) +\frac{(m-n)^2 X_0^2}{(1+X_0^2)} \right]\mathcal{V}(\rho h^{(m)}_{00})
\nonumber \\-\frac{4 G}{9} \left[  2 m n
  (1+X_0^2) -\frac{(m-n)^2 X_0^2}{(1+X_0^2)}\right]\mathcal{V}( \rho h^{(m)}_{ii})
\nonumber\\
+\frac{8 G (m+nX_0^2) (m-n)}{3}  \left[U+\Phi_3- 3 \Phi_4
  + \frac{2 }{(1+X_0^2)}\mathcal{V}(\rho h^{(0)}_{00})  \right]
\nonumber \\
 -\frac{(2 \ka_1+\ka_2)}{3 \pi X_0^2 l^2}
\Bigg[
2 \mathcal{V}(h^{(0)}_{00} A)
+8  \mathcal{V}(h^{(0)}_{00} h^{(m)}_{00})
-4 \mathcal{V}(h^{(m)2}_{00})
+ \mathcal{V}(A^2)
-4 \mathcal{V}(h^{(m)2}_{ij})
\Bigg] \nonumber \\
-\frac{ \left[ \mathcal{V}(h^{(0)}_{00}
  \nabla^2 h^{(m)}_{ii})+\mathcal{V}(h^{(m)}_{ii}\nabla^2 h^{(0)}_{00}) +2
  \mathcal{V}(h^{(0)}_{00,i} h^{(m)}_{jj,i})
  \right]}{4\pi (1+X_0^2)}
+\frac{ \mathcal{V}( h^{(m)}_{00,00} ) }{2 \pi}
\nonumber \\-\frac{\mathcal{V}(h^{(0)}_{00,i} h^{(m)}_{00,i})}{6 \pi (1+X_0^2)}
+\frac{ \left[ 8 \mathcal{V}(h^{(m)}_{ij}
  h^{(0)}_{00,ij})+5\mathcal{V}(h^{(0)}_{00,i} A_{,i})
  \right]}{24 \pi (1+X_0^2)}
-\frac{ \mathcal{V}(A_{,00})}{4\pi}
\nonumber \\
+\frac{ \left[ \mathcal{V}(h^{(0)}_{00}
  \nabla^2 h^{(m)}_{00})+\mathcal{V}(h^{(m)}_{00} \nabla^2
  h^{(0)}_{00}) \right]}{12 \pi (1+X_0^2)}
 -\frac{(4\ka_1+3 (1-X_0^2) \ka_2)}{4 \pi M^2 l^2 X_0^2}
\mathcal{V}( A_{,i}^2)
\nonumber \\
+\frac{(3(1-X_0^2) \ka_1+(1-2X_0^2) \ka_2)}{6 \pi (1+X_0^2) (2\ka_1+\ka_2)}
\mathcal{V}(h^{(m)2}_{ij,k})
+\frac{ \mathcal{V}(h^{(m)}_{ij} A_{,ij})}{6\pi (1+X_0^2)}
\nonumber \\+\frac{(4 (1-X_0^2) \ka_1+(1-3X_0^2) \ka_2)}{12 \pi (1+X_0^2) (2\ka_1+\ka_2)}
\mathcal{V}( h^{(m)}_{ij} \nabla^2
h^{(m)}_{ij} )-\frac{\ka_1 \mathcal{V}(A \nabla^2 A)}{12 \pi(2\ka_1+\ka_2)}
\nonumber\\+\frac{X_0^2}{12 \pi(1+X_0^2)} \left[\mathcal{V}(A_{,i} h^{(m)}_{00,i})
  +4 \mathcal{V}(h^{(m)}_{ij} h^{(m)}_{00,ij})\right]
+ \frac{(4 \ka_1+ \ka_2)}{12 \pi(2\ka_1+\ka_2)}
\mathcal{V}(h^{(m)}_{00}
  \nabla^2 h^{(m)}_{00})
\nonumber \\ + \frac{((3+X_0^2)\ka_1+(1+3X_0^2) \ka_2)}{6 \pi(1+X_0^2)(2\ka_1+\ka_2)}
\mathcal{V}(h^{(m)2}_{00,i})
+ \frac{(\ka_1 + \ka_2)}{3 \pi (2\ka_1+\ka_2)}
\mathcal{V}(h^{(m)}_{ij,k} h^{(m)}_{jk,i}).
\label{A42}
\eea

\vspace{15pt}
\noindent
{\bf Other theories}
\vspace{15pt}
\bea
(X_0^2 \ka_0+12 \ka_1+6\ka_2) A \nonumber \\
\hspace{-15pt} =
\frac{3 M^2 l^2 X_0^2}{8\pi (1+X_0^2)} \mathcal{W}_N( h^{(m)}_{00,00} )
+\frac{G M^2 l^2 X_0^2}{(1+X_0^2)} \left[ 3 m n (1-X_0^2) +\frac{(m-n)^2 X_0^2}{(1+X_0^2)} \right]\mathcal{W}_N(\rho h^{(m)}_{00})
\nonumber\\
-\frac{3 (2\ka_1+\ka_2)}{2\pi} \mathcal{W}_N(A_{,00})
-\frac{G M^2 l^2 X_0^2}{3 (1+X_0^2)} \left[  2 m n
  (1+X_0^2) -\frac{(m-n)^2 X_0^2}{(1+X_0^2)}\right]\mathcal{W}_N( \rho h^{(m)}_{ii})
\nonumber\\
+\frac{2 G M^2 l^2 X_0^2
(m+nX_0^2) (m-n)}{(1+X_0^2)}  \Big[\mathcal{W}_N(\rho)+\mathcal{W}_N(\rho \Pi)- 3 \mathcal{W}_N(p)
 \nonumber \\ \qquad \qquad \qquad \qquad\qquad \qquad \qquad\qquad \qquad  \qquad + \frac{2 }{(1+X_0^2)}\mathcal{W}_N(\rho h^{(0)}_{00})  \Big]
\nonumber \\ -\frac{M^2 (2 \ka_1+\ka_2)}{4\pi (1+X_0^2)}
\Big[
2 \mathcal{W}_N(h^{(0)}_{00} A)
+8  \mathcal{W}_N(h^{(0)}_{00} h^{(m)}_{00})
\nonumber \\ \qquad \qquad \qquad \qquad \qquad-4 \mathcal{W}_N(h^{(m)2}_{00})
+ \mathcal{W}_N(A^2)
-4 \mathcal{W}_N(h^{(m)2}_{ij})
\Big] \nonumber \\
-\frac{(X_0^2 \ka_0+12\ka_2+6\ka_2)}{4\pi (1+X_0^2)} \left[ \mathcal{W}_N(h^{(0)}_{00}
  \nabla^2 h^{(m)}_{ii})+\mathcal{W}_N(h^{(m)}_{ii}\nabla^2 h^{(0)}_{00}) +2
  \mathcal{W}_N(h^{(0)}_{00,i} h^{(m)}_{jj,i})
  \right]\nonumber \\ -\frac{(2\ka_1+\ka_2)}{\pi (1+X_0^2)}
\mathcal{W}_N(h^{(0)}_{00,i} h^{(m)}_{00,i})
+\frac{M^2 l^2 X_0^2}{8\pi (1+X_0^2)^2} \left[ 2 \mathcal{W}_N(h^{(m)}_{ij}
  h^{(0)}_{00,ij})+5\mathcal{W}_N(h^{(0)}_{00,i} h^{(m)}_{ij,j})  \right]
\nonumber \\
+\frac{(X_0^2\ka_0+4\ka_1+2\ka_2)}{4 \pi (1+X_0^2)} \left[ \mathcal{W}_N(h^{(0)}_{00}
  \nabla^2 h^{(m)}_{00})+\mathcal{W}_N(h^{(m)}_{00} \nabla^2 h^{(0)}_{00}) \right]
\nonumber \\
-\frac{(X_0^2 \ka_0+2 (1+X_0^2) \ka_2)}{4 \pi (1+X_0^2)}
\left[ \mathcal{W}_N(h^{(m)}_{ij,k}h^{(m)}_{ij,k}) +\mathcal{W}_N( h^{(m)}_{ij} \nabla^2
h^{(m)}_{ij} )\right]
\nonumber \\
 -\frac{(X_0^2 \ka_0 +4 (1-X_0^2) \ka_1 +4 \ka_2)}{4 \pi (1+X_0^2)}
\left[\mathcal{W}_N( h^{(m)}_{ij,i} A_{,j})+ \mathcal{W}_N(h^{(m)}_{ij} A_{,ij})\right]
\nonumber\\ +\frac{M^2l^2 X_0^4}{4\pi(1+X_0^2)^2} \left[\mathcal{W}_N(h^{(m)}_{ij,i} h^{(m)}_{00,j})
  +\mathcal{W}_N(h^{(m)}_{ij} h^{(m)}_{00,ij})\right]
\nonumber \\
+\frac{(X_0^2 \ka_0+2 \ka_2)}{8\pi} \left[\mathcal{W}_N(A_{,i}^2)
  +\mathcal{W}_N(A \nabla^2 A)-2 \mathcal{W}_N(h^{(m)2}_{00,i})-2 \mathcal{W}_N(h^{(m)}_{00}
  \nabla^2 h^{(m)}_{00} )\right]
\nonumber \\
-\frac{(X_0^2\ka_0 +4\ka_1 +4 \ka_2)}{4\pi} \left[ \mathcal{W}_N(A_{,i}
  h^{(m)}_{ij,j})+\mathcal{W}_N(A h^{(m)}_{ij,ij})\right]
\nonumber \\
+ \frac{X_0^2 (3 X_0^2 \ka_0+16 \ka_1 +8 \ka_2 )}{4\pi(1+X_0^2)}
\left[ \mathcal{W}_N(h^{(m)2}_{00,i})+\mathcal{W}_N(h^{(m)}_{00}
  \nabla^2 h^{(m)}_{00}) \right]
\nonumber \\
+ \frac{(X_0^2 \ka_0 +8 \ka_1 +2 (3+X_0^2) \ka_2)}{2 \pi (1+X_0^2)} \left[
  \mathcal{W}_N(h^{(m)2}_{ij,j})+  2 \mathcal{W}_N(h^{(m)}_{ij}
  h^{(m)}_{jk,ki})+\mathcal{W}_N(h^{(m)}_{ij,k} h^{(m)}_{jk,i})\right]
\nonumber \\
  +\frac{M^2 l^2 X_0^2(1-X_0^2)}{16\pi(1+X_0^2)^2} \Bigg[ 3\mathcal{W}_N(h^{(m)2}_{00,i}) +
  2 \mathcal{W}_N(h^{(m)}_{00} \nabla^2 h^{(m)}_{00}) - \mathcal{W}_N(A_{,i}^2)
  -2 \mathcal{W}_N(h^{(m)}_{ij,k}h^{(m)}_{ik,j})
 \nonumber\\ \qquad \qquad  \qquad \qquad
+2 \mathcal{W}_N(h^{(m)}_{jk}\nabla^2h^{(m)}_{jk})
  +3 \mathcal{W}_N(h^{(m)2}_{ij,k}) - 4 \mathcal{W}_N(h^{(m)2}_{ij,j})
\nonumber\\ \qquad \qquad  \qquad \qquad
+ 4 \mathcal{W}_N(h^{(m)}_{ij,j}
  A_{,i}) +2 \mathcal{W}_N(h^{(m)}_{ij} A_{,ij}) -
  4 \mathcal{W}_N(h^{(m)}_{jk} h^{(m)}_{ij,ik})\Bigg].
\label{A43}
\eea
In the above we have made use of the Bianchi identities to $O(4)$, which are given by:
\bea
 (X_0^2 \ka_0+8\ka_1+4\ka_2) h^{(m)}_{ij,ij} \nonumber \\
\hspace{-15pt} = \frac{1}{2} (X_0^2 \ka_0+4 \ka_1+2 \ka_2) \nabla^2A +\frac{(X_0^2\ka_0 +4\ka_1 +4 \ka_2)}{2} \left[ A_{,i}
  h^{(m)}_{ij,j}+A h^{(m)}_{ij,ij}\right]
\nonumber\\
 +\frac{(X_0^2 \ka_0+2 (1+X_0^2) \ka_2)}{2 (1+X_0^2)}
\left[ h^{(m)}_{ij,k}h^{(m)}_{ij,k} + h^{(m)}_{ij} \nabla^2
h^{(m)}_{ij} \right]
\nonumber \\
-\frac{(X_0^2 \ka_0+2 \ka_2)}{4} \left[A_{i} h^{(m)}_{jj,i} +A
  \nabla^2 h^{(m)}_{ii}
-h^{(m)}_{jj,i}
  h^{(m)}_{00,i} - h^{(m)}_{ii} \nabla^2 h^{(m)}_{00}
\right]\nonumber \\- \frac{(X_0^2 \ka_0 +4
  \ka_1+2 \ka_2)}{2 (1+X_0^2)} \left[h^{(0)}_{00,i}
  h^{(m)}_{jj,i}+h^{(0)}_{00} \nabla^2 h^{(m)}_{ii}
+h^{(0)}_{00,i} h^{(m)}_{00,i}+  h^{(0)}_{00} \nabla^2 h^{(m)}_{00}
\right] \nonumber
\\ +\frac{(X_0^2 \ka_0 +4 (1-X_0^2) \ka_1 +4 \ka_2)}{2 (1+X_0^2)}
\left[ h^{(m)}_{ij,i} h^{(m)}_{kk,j}+ h^{(m)}_{ij} h^{(m)}_{kk,ij}
  \right] \nonumber \\
 -\frac{(X_0^2 \ka_0 +6 \ka_1+3 \ka_2)}{(1+X_0^2)} \left[
  h^{(m)}_{00,i} h^{(0)}_{00,i}+  h^{(m)}_{00} \nabla^2 h^{(0)}_{00}
  \right] \nonumber \\+ \frac{(2\ka_1+ \ka_2)}{(1+X_0^2)} \left[ h^{(m)}_{jj,i}
  h^{(0)}_{00,i} + h^{(m)}_{ii} \nabla^2 h^{(0)}_{00} \right]
 - (X_0^2
\ka_0+8 \ka_1+4 \ka_2) h^{(m)}_{0i,0i}
\nonumber \\ + \frac{(X_0^2 (1-5X_0^2) \ka_0-32 X_0^2 \ka_1 +2
  (1-7X_0^2) \ka_2 )}{4 (1+X_0^2)} \left[ h^{(m)2}_{00,i}+h^{(m)}_{00}
  \nabla^2 h^{(m)}_{00} \right] \nonumber \\
- \frac{(X_0^2 \ka_0 +8 \ka_1 +2 (3+X_0^2) \ka_2)}{(1+X_0^2)} \left[
  h^{(m)}_{ij,i} h^{(m)}_{jk,k}+  2 h^{(m)}_{ij}
  h^{(m)}_{jk,ki}+h^{(m)}_{ij,k} h^{(m)}_{jk,i}
  \right] \nonumber \\
+\frac{(X_0^2 \ka_0+8 \ka_1+4 \ka_2)}{(1+X_0^2)} \left[h^{(0)}_{00}
  h^{(m)}_{ij,ij} - h^{(m)}_{ij} h^{(0)}_{00,ij} \right]
\nonumber \\
-\frac{((1+2X_0^2) (X_0^2 \ka_0+4 \ka_2) +4 (1+3X_0^2) \ka_1)}{2
  (1+X_0^2)} \left[h^{(m)}_{ij,i} h^{(m)}_{00,j} +h^{(m)}_{ij}
  h^{(m)}_{00,ij} \right].
\eea

\newpage
\section*{Appendix C: The {\boldmath $\hat{h}_{\mu \nu}$}
  Equations, and Gauge Transformations}

As long as $m\neq -nX_0^2$ we can always use a coordinate transformation to
remove all contributions from the $O(1)$ massive modes to $\hat{h}_{0i}$, so that $\hat{h}_{0i} \rightarrow 0$.
We will therefore consider here only the higher order
contributions to $\hat{h}_{\mu \nu}$.

\subsection*{{\bf Newtonian perturbations to {\boldmath $O(2)$}}}

To $O(2)$ in $\hat{h}_{00}$ we find the results below.

\bigskip
\noindent
{\bf ISS theory}
\bea
\hspace{-30pt}
\hat{h}_{00} = 2 G_N U -
\frac{\sigma (m+n \sigma)}{4 \pi (1+\sigma)^2} \left[
  \mathcal{V}\left(h^{(m)2}_{0i,j}\right)-\mathcal{V}\left(h^{(m)}_{0i,j}h^{(m)}_{0j,i}\right) \right]
\nonumber \\ + \frac{(11-9 \sigma) (m-n) \sigma}{24 \pi (1+\sigma)^2} \mathcal{W}_M
\left(h^{(m)2}_{0i,j}\right)+\frac{8 G_N (m-n)^2 \sigma}{3(m+n \sigma)^2}  \mathcal{W}_M \left(\rho\right)
\nonumber \\
 - \frac{(1-3 \sigma)(m-n) \sigma}{24 \pi (1+\sigma)^2} \mathcal{W}_M
 \left(h^{(m)}_{0i,j} h^{(m)}_{0j,i}\right)
 -\frac{\sigma (m+n \sigma)}{2\pi l^2
  (1+\sigma)}\mathcal{V}\left(h^{(m)2}_{0i}\right)
\nonumber \\ + \frac{(1-4 \sigma)(m-n)\sigma}{6 \pi l^2 (1+\sigma)} \mathcal{W}_M
\left(h^{(m)2}_{0i}\right).
\eea

\bigskip
\noindent
{\bf $\mathbf{\ka_0=0}$ theory}
\bea
\hspace{-30pt}
\hat{h}_{00} =   2\mathcal{G}_N U
+\frac{8 (m-n)^2 X_0^2 \mathcal{G}_N}{[ 3 (m+n
  X_0^2)^2 -X_0^2 (m-n)^2 ]} \mathcal{W}_M\left(\rho\right)
\nonumber \\
- \frac{X_0^2 [(m-n) ((X_0^2-3) \ka_1+(X_0^2-1) \ka_2)+3 (m+nX_0^2) (2 \ka_1+\ka_2)]}{12 \pi (1+X_0^2)^2 (2\ka_1+\ka_2)}
\mathcal{V}\left(h^{(m)2}_{0i,j}\right)
\nonumber \\
+\frac{X_0^2[(m-n) ((1-X_0^2) \ka_1+\ka_2) +3 (m+nX_0^2) (2\ka_1+\ka_2)]}{12 \pi (1+X_0^2)^2 (2\ka_1+\ka_2)}
\mathcal{V}\left(h^{(m)}_{0i,j} h^{(m)}_{0j,i}\right)
\nonumber \\
+ \frac{[(m-n) (8\ka_1+(3-X_0^2) \ka_2)-3 (m+nX_0^2)(2\ka_1+\ka_2)]}{3 \pi l^2 (1+X_0^2)}
\mathcal{V}\left(h^{(m)2}_{0i}\right)
\nonumber \\
+ \frac{(1-X_0^2)(m-n) X_0^2}{4
  \pi(1+X_0^2)^2}\left[\mathcal{W}_M\left(h^{(m)2}_{0i,j}\right)-\mathcal{W}_M\left(h^{(m)}_{0i,j}h^{(m)}_{0j,i}\right)\right]
\nonumber \\ + \frac{(4 (1-X_0^2)\ka_1+(1-3X_0^2) \ka_2)(m-n)}{2\pi l^2 (1+X_0^2)}
\mathcal{W}_M\left(h^{(m)2}_{0i}\right).
\eea

\bigskip
\noindent
{\bf Other theories}
%
\bea
\hspace{-30pt}
\hat{h}_{00} = 2 G_N U - \frac{X_0^2 (m+n X_0^2)}{4 \pi (1+X_0^2)^2} \left[
  \mathcal{V}\left(h^{(m)2}_{0i,j}\right)-\mathcal{V}\left(h^{(m)}_{0i,j}h^{(m)}_{0j,i}\right) \right]
\nonumber \\ -\frac{M^2X_0^2(m+n X_0^2)}{8\pi
  (1+X_0^2)^2}\mathcal{V}\left(h^{(m)2}_{0i}\right)
+\frac{(m-n)(8\ka_1+(3-X_0^2) \ka_2)}{3 \pi l^2 (1+X_0^2)} \mathcal{W}_N\left(h^{(m)2}_{0i}\right)
\nonumber \\
+   \frac{8 G_N X_0^2 (m-n)^2}{3(m+n X_0^2)^2}
\mathcal{W}_M\left(\rho\right)
- \frac{2 G_N X_0^2(m-n)^2}{3 (m+n X_0^2)^2} \mathcal{W}_N\left(\rho \right)
\nonumber \\ + \frac{(1-X_0^2)(m-n) X_0^2}{4
  \pi(1+X_0^2)^2}\left[\mathcal{W}_M\left(h^{(m)2}_{0i,j}\right)-\mathcal{W}_M\left(h^{(m)}_{0i,j}h^{(m)}_{0j,i}\right)\right]
\nonumber \\ -\frac{(m-n)(X_0^4 \ka_0-8 (1-X_0^2) \ka_1-2 (1-3X_0^2) \ka_2)}{4\pi
  l^2 (1+X_0^2)} \mathcal{W}_M\left(h^{(m)2}_{0i}\right)
\nonumber \\\nonumber + \frac{(m-n)(X_0^2 (1-X_0^2)\ka_0 +8 (3-X_0^2) \ka_1 +8
  (1-X_0^2)\ka_2)}{12\pi M^2l^2 (1+X_0^2)} \mathcal{W}_N\left(h^{(m)2}_{0i,j}\right)
\\ +\frac{(m-n)(X_0^2 (1-X_0^2) \ka_0+8(1-X_0^2) \ka_1
  +8\ka_2)}{12\pi M^2l^2 (1+X_0^2)}
\mathcal{W}_N\left(h^{(m)}_{0i,j} h^{(m)}_{0j,i}\right).
\eea
%

\subsection*{{\bf Post-Newtonian perturbations to {\boldmath $O(2)$} and {\boldmath $O(3)$}}}

To $O(2)$ in $\hat{h}_{ij}$ and $O(3)$ in $\hat{h}_{0i}$ we find the
results below.

\bigskip
\noindent
{\bf ISS theory}
%
\bea
\hspace{-30pt}
\hat{h}_{ij} = 2 G_N \delta_{ij} U -\frac{l^2 G_N \sigma (m-n)^2}{3 (1+\sigma)(m+n \sigma)^2}
\mathcal{W}_M(\rho_{,ij}) +\frac{4 G_N \sigma (m-n)^2}{3 (m+n
  \sigma)^2} \delta_{ij} \mathcal{W}_M(\rho)\\
\hspace{-15pt}\rightarrow  2 G_N \delta_{ij} U+\frac{4 G_N \sigma (m-n)^2}{3 (m+n
  \sigma)^2} \delta_{ij} \mathcal{W}_M(\rho)
\eea
and
\bea
\hspace{-30pt}
\hat{h}_{0i} = -\frac{7G_N}{2} V_i - \frac{G_N}{2} W_i
- \frac{4 G_N (m-n)^2 \sigma}{(m+n \sigma)^2} \mathcal{W}_M (\rho
v_i)
\nonumber \\  -\frac{G_N l^2 \sigma (m-n)^2}{3 (1+\sigma)(m+n
  \sigma)^2} \mathcal{W}_M(\rho_{,0i}) \\
\hspace{-15pt}\rightarrow -\frac{7G_N}{2} V_i - \frac{G_N}{2} W_i
- \frac{4 G_N (m-n)^2 \sigma}{(m+n \sigma)^2} \mathcal{W}_M (\rho
v_i).
\eea
The arrows in the expressions indicate the infinitesimal coordinate transformations
\bea
\xi_i = -\frac{G_N l^2 \sigma (m-n)^2}{6 (1+\sigma)(m+n\sigma)^2}
\mathcal{W}_M(\rho)_{,i}\\
\xi_0 = -\frac{G_N l^2 \sigma (m-n)^2}{6 (1+\sigma)(m+n\sigma)^2}
\mathcal{W}_M(\rho)_{,0}.
\eea

\bigskip
\noindent
{\bf $\mathbf{\ka_0=0}$ theory}
%
\bea
\hspace{-30pt}
\hat{h}_{ij} = 2 \mathcal{G}_N \frac{\left( 3(m+n
X_0^2)^2  +X_0^2 (m-n)^2  \right)}{\left( 3(m+n
X_0^2)^2  -X_0^2 (m-n)^2  \right)} \delta_{ij} U
\nonumber \\+\frac{4 \mathcal{G}_N X_0^2 (m-n)^2}{\left( 3(m+n
X_0^2)^2  -X_0^2 (m-n)^2  \right)}  \delta_{ij} \mathcal{W}_M(\rho)
\nonumber \\+\frac{\mathcal{G}_N X_0^2 (m-n)^2}{\pi \left( 3(m+n
X_0^2)^2  -X_0^2 (m-n)^2  \right)}  \mathcal{W}_M(U_{,ij})\\
\hspace{-15pt}\rightarrow  2 \mathcal{G}_N \frac{\left( 3(m+n
X_0^2)^2  +X_0^2 (m-n)^2  \right)}{\left( 3(m+n
X_0^2)^2  -X_0^2 (m-n)^2  \right)} \delta_{ij} U
\nonumber \\+\frac{4 \mathcal{G}_N X_0^2 (m-n)^2 \delta_{ij}}{\left( 3(m+n
X_0^2)^2  -X_0^2 (m-n)^2  \right)}  \mathcal{W}_M(\rho)
\eea
and
\bea
\hspace{-30pt} \hat{h}_{0i} =
 - \frac{4 G X_0^2 (m+n X_0^2)(m-n)^2}{(1+X_0^2)}
\mathcal{W}_M (\rho v_i) -\frac{7G (m+n X_0^2)^3}{2(1+X_0^2)} V_i
\nonumber \\+ \frac{Gl^2 X_0^4 (m+nX_0^2)(m-n)^2}{6 (1+X_0^2)^2 (2 \ka_1+\ka_2)}
U_{,0i} - \frac{G
  (m+n X_0^2)^3}{2(1+X_0^2)} W_i \nonumber \\ - \frac{Gl^2 X_0^4 (m+nX_0^2) (m-n)^2}{6 (1+X_0^2)^2
  (2\ka_1+\ka_2)} \mathcal{W}_M(\rho_{,0i})
\\ \hspace{-15pt} \rightarrow -\frac{7G (m+n X_0^2)^3}{2(1+X_0^2)} V_i - \frac{G
  (m+n X_0^2)^3}{2(1+X_0^2)} W_i \nonumber \\- \frac{4 G X_0^2 (m+n X_0^2)(m-n)^2}{(1+X_0^2)}
\mathcal{W}_M (\rho v_i),
\eea
where the arrows indicate the coordinate transformations
\bea
\hspace{-40pt}
\xi_i = \frac{\mathcal{G}_N l^2 X_0^4 (m-n)^2}{4 (1+X_0^2)^2 (2
  \ka_1+\ka_2) (3(m+nX_0^2)^2-X_0^2(m-n)^2)} \mathcal{W}_M(\rho)_{,i}\\
\hspace{-40pt} \xi_0 = \frac{\mathcal{G}_N l^2 X_0^4 (m-n)^2}{4 (1+X_0^2)^2 (2
  \ka_1+\ka_2) (3(m+nX_0^2)^2-X_0^2(m-n)^2)} (2 U_{,0}-3 \mathcal{W}_M(\rho)_{,0}).
\eea

\bigskip
\noindent
{\bf Other theories}
%
\bea
\hspace{-30pt}
\hat{h}_{ij} = 2 G_N \delta_{ij} U
+ \frac{4 G_N X_0^2 (m-n)^2}{3(m+n X_0^2)^2 } \delta_{ij} \mathcal{W}_M(\rho)
+\frac{2 G_N X_0^2 (m-n)^2}{3(m+n X_0^2)^2 }  \delta_{ij} \mathcal{W}_N(\rho)
\nonumber \\
+ \frac{4 G_N X_0^2 (m-n)^2}{3 M^2 (m+n X_0^2)^2} (\mathcal{W}_N(\rho_{,ij}) -
\mathcal{W}_M(\rho_{,ij}))\\
\hspace{-15pt} \rightarrow  2 G_N \delta_{ij} U
+ \frac{4 G_N X_0^2 (m-n)^2}{3(m+n X_0^2)^2 } \delta_{ij} \mathcal{W}_M(\rho)
+\frac{2 G_N X_0^2 (m-n)^2}{3(m+n X_0^2)^2 }  \delta_{ij}
\mathcal{W}_N(\rho)
\eea
and
\bea
\hspace{-30pt}
\hat{h}_{0i} = -\frac{7G_N}{2} V_i - \frac{G_N}{2} W_i
- \frac{4 G_N X_0^2 (m-n)^2}{(m+n X_0^2)^2}\mathcal{W}_M (\rho v_i)
\nonumber \\
+\frac{4 G_N X_0^2 (m-n)^2}{3 M^2 (m+nX_0^2)^2 } (
\mathcal{W}_N(\rho_{,0i})-\mathcal{W}_M(\rho_{,0i}))
\\ \hspace{-15pt} \rightarrow -\frac{7G_N}{2} V_i - \frac{G_N}{2} W_i
- \frac{4 G_N X_0^2 (m-n)^2}{(m+n X_0^2)^2}\mathcal{W}_M (\rho v_i)
\eea
where the arrows indicate the coordinate transformations
\bea
\xi_i = \frac{2 G_N X_0^2 (m-n)^2}{3 M^2 (m+nX_0^2)^2 } (
\mathcal{W}_N(\rho)-\mathcal{W}_M(\rho))_{,i}
\\
\xi_0 = \frac{2 G_N X_0^2 (m-n)^2}{3 M^2 (m+nX_0^2)^2 } (
\mathcal{W}_N(\rho)-\mathcal{W}_M(\rho))_{,0}.
\eea
We will not show the full expressions for $\hat{h}_{00}$ to $O(4)$
here as they are quite lengthy, and we do not feel that writing them
out explicitly will add sufficient extra insight to justify their
inclusion.  The interested reader can calculate these quantities
straightforwardly using the relevant expressions in Appendix B, and
the coordinate transformations given above.

\end{document}